\documentclass[sigconf]{acmart}

\AtBeginDocument{%
  }
\setcopyright{acmlicensed}
\copyrightyear{}
\acmYear{2026}
\acmDOI{}
\acmConference[CCS'26]{}{November 15--19, 2026}{The Hague, The Netherlands}

\usepackage{mathtools}
\usepackage{subcaption}
\usepackage{rotating}
\usepackage{colortbl}
\usepackage{pgfplots,pgfplotstable}
\pgfplotsset{compat=1.18}
\usepackage{booktabs}
\usepackage{float}
\usepackage{boxedminipage}
\usepackage{enumitem}
\usepackage{framed}
\usepackage{esvect}
\usepackage{tikz}
\usepackage{latexsym}
\usepackage{blkarray}
\usepackage{prettyref}
\usepackage{multirow}
\usepackage{tablefootnote}
\usepackage{longtable}
\usepackage{xspace}
\usepackage{ulem}
\usepackage{array}
\usepackage{todonotes}
\usepackage[most]{tcolorbox}
\usepackage{mdframed}

\definecolor{col3}{HTML}{1f77b4}
\definecolor{col5}{HTML}{ff7f0e}
\definecolor{col7}{HTML}{1d3ada}
\definecolor{col8}{HTML}{1f2fc1}
\definecolor{col10}{HTML}{2ca02c}
\definecolor{colS128}{HTML}{9467bd}
\definecolor{colS256}{HTML}{d62728}
\definecolor{colC3072}{HTML}{8c564b}
\definecolor{lightgray}{gray}{0.92}
\pgfplotsset{
  ccs/.style={
    width=\linewidth,
    height=5.5cm,
    grid=both,
    grid style={line width=0.3pt, draw=gray!30},
    tick label style={font=\small},
    xticklabel style={rotate=45, anchor=east, font=\small},
    label style={font=\small},
    legend style={font=\footnotesize, draw=black!40,
                  fill=white, inner sep=2pt},
    every axis plot/.append style={line width=1.2pt},
    enlarge x limits=0.04,
    enlarge y limits=0.06,
  }
}

\definecolor{reviewyellow}{RGB}{255,215,0}
\newmdenv[
    leftline=true,
    rightline=false,
    topline=false,
    bottomline=false,
    linewidth=4pt,
    linecolor=reviewyellow,
    skipabove=\baselineskip,
    skipbelow=\baselineskip,
    innerleftmargin=10pt,
    innerrightmargin=0pt,
    innertopmargin=6pt,
    innerbottommargin=6pt
]{changesblock}

\newtcolorbox{mybox}[2][]{%
  attach boxed title to top center={yshift=-10pt},
  colframe=black,
  colbacktitle=black,
  title=#2,#1,
  enhanced,
}

\newtheorem{assumption}{Assumption}

\newcommand{\lnum}{\refstepcounter{equation}{\st\theequation:}\hspace{.8mm}}
\newcommand{\countt}{\refstepcounter{equation}{\theequation}}

\usepackage{algorithm}
\usepackage{algpseudocode}

\usepackage{arydshln}

\usepackage{enumitem}

\usepackage[dvipsnames, svgnames]{xcolor}

\usepackage{float} 
\usepackage{amsmath} 
\newlist{steps}{enumerate}{1}
\setlist[steps, 1]{label = Case \arabic*}

\usepackage{algpseudocode}

\usepackage{array}
\usepackage{multirow}

\newcommand{\hs}{\hspace{-5.6mm}}

\newcommand{\inptCln}{\ensuremath{\textsf{Inp}}\xspace}

\newcommand{\oracle}{\ensuremath{\textsf{Oracle}}\xspace}

\newcommand{\subprt}{\ensuremath{_{u}}\xspace}

\newcommand{\adv}{\ensuremath{\mathcal{A}}\xspace}

\newcommand{\chal}{\ensuremath{\mathcal{E}}\xspace}

\newcommand{\advb}{\ensuremath{\prtt}\xspace}

\newcommand{\corupt}{\ensuremath{\mathcal{W}}\xspace}

\newcommand{\et}{\textit{et al}\xspace}

\newcommand{\cor}{\ensuremath{\pt}\xspace}

\newcommand{\rt}{\ensuremath{root}\xspace}
\newcommand{\vhtlp}{\ensuremath{\mathsf{VHLC\text{-}TLP}}\xspace}

\newcommand{\tf}{Tempora-Fusion\xspace}

\newcommand{\tl}{\ensuremath{\ell}\xspace}

\newcommand{\kp}{\ensuremath{K}\xspace}
\newcommand{\cmd}{\ensuremath {cmd}\xspace}

\newcommand{\valSetup}{\ensuremath{\mathsf{ValidateSetup}}\xspace}

\newcommand{\mxsqr}{\ensuremath{max_{\st ss}}\xspace}

\newcommand{\mxs}{\ensuremath{max_{\st cs}}\xspace}

\newcommand{\srv}{\ensuremath {s}\xspace}
\newcommand{\cln}{\ensuremath {c}\xspace}

\newcommand{\prm}{\ensuremath{{p}}\xspace}

\newcommand{\ssetup}{\ensuremath{\mathsf{S.Setup}}\xspace}
\newcommand{\csetup}{\ensuremath{\mathsf{C.Setup}}\xspace}
\newcommand{\pgen}{\ensuremath{\mathsf{\mathsf{GenPuzzle}}}\xspace}
\newcommand{\eval}{\ensuremath{\mathsf{\mathsf{Eval}}}\xspace}
\newcommand{\solv}{\ensuremath{\mathsf{\mathsf{Solve}}}\xspace}
\newcommand{\ver}{\ensuremath{\mathsf{\mathsf{Verify}}}\xspace}

\newcommand{\asn}{\ensuremath{\coloneqq}\xspace}

\newcommand{\modi}{\ensuremath{\mathcal{Q}}\xspace}

\newcommand{\enco}{\ensuremath{u}\xspace}

\newcommand{\pt}{\ensuremath{\alpha}\xspace}

\newcommand{\sample}{\ensuremath{\stackrel{\st\$}\leftarrow}\xspace}

\newcommand{\F}{\ensuremath{\mathbb{F}}\xspace}

\newcommand{\roott}{\ensuremath{\beta}\xspace}

\newcommand{\opgen}{\ensuremath{\mathsf{\mathsf{O.Gp}}}\xspace}
\newcommand{\oeval}{\ensuremath{\mathsf{\mathsf{O.Ev}}}\xspace}
\newcommand{\orev}{\ensuremath{\mathsf{\mathsf{O.Rv}}}\xspace}

\newcommand{\scp}{\ensuremath{\text{cltPzl}}\xspace}

\newcommand{\ep}{\ensuremath{\text{evlPzl}}\xspace}

\newcommand{\cl}{\ensuremath{\mathrm c}\xspace}

\newcommand{\prt}{\ensuremath{u }\xspace}

\newcommand{\prtt}{\ensuremath{c}\xspace}

\newcommand{\idx}{\ensuremath{\mathcal{I} }\xspace}

\newcommand{\ole}{\ensuremath{\mathtt{OLE}}\xspace}

\newcommand{\prf}{\ensuremath{\mathtt{PRF} }\xspace}

\newcommand{\g}{\ensuremath{\mathtt{G} }\xspace}

\newcommand{\pnp}{\ensuremath{ \phi(N_{\st \prt}) }\xspace}

\newcommand{\set}{\ensuremath{\{\cln_{\st 1},\ldots, \cln_{\st n}\}}\xspace}

\newcommand{\comcom}{\ensuremath{\mathtt{Com}}\xspace}
\newcommand{\comver}{\ensuremath{\mathtt{Ver}}\xspace}

\newcommand{\st}{\scriptscriptstyle}

\newcommand{\func}{\ensuremath{\mathcal{F}^{\st \text{PLC}}}\xspace}

\DeclareRobustCommand{\bm}[1]{\boldsymbol{#1}} 

\let\emph\relax
\DeclareTextFontCommand{\emph}{\itshape}



\begin{document}

\title{\tf: Time-Lock Puzzle with Efficient Verifiable\\ Homomorphic Linear Combination}

\author{Aydin Abadi}
\orcid{0000-0002-1414-8351}
\authornote{Corresponding author}
\affiliation{%
    \department[0]{Cryptography and AI Security Lab}
  \department[1]{School of Computing}
  \institution{Newcastle University}
  \city{Newcastle upon Tyne}
  \country{United Kingdom}
}
\email{aydin.abadi@ncl.ac.uk}

\author{Jakub K. Szelag}
\orcid{0009-0000-9635-0598}
\affiliation{%
    \department[0]{Cryptography and AI Security Lab}
  \department[1]{School of Computing}
  \institution{Newcastle University}
  \city{Newcastle upon Tyne}
  \country{United Kingdom}
}
\email{J.K.Szelag2@newcastle.ac.uk}


\begin{abstract}
Homomorphic time-lock puzzles (TLPs) let parties lock sensitive values now so that they become recoverable only after a designated delay, without requiring the owners to remain available at release time. In many settings, however, an authorized aggregate over several hidden inputs may need to be released and publicly verified before the underlying inputs themselves are opened. Existing homomorphic TLPs support computation over puzzles; however, these TLPs do not offer an efficient mechanism for publicly verifying that a released result is the prescribed linear combination of the intended puzzles. Homomorphic timed commitments provide publicly recomputable aggregation and efficiently verifiable forced opening, but existing constructions place the commitments and their aggregate under a common delay. Thus, they cannot give the aggregate a shorter, independently chosen release time while the constituent values remain locked under their respective delays.

We present \tf, the first homomorphic TLP scheme with efficient public verification of both individual puzzle solutions and homomorphic linear combinations. \tf lets clients generate puzzles independently, later authorize a linear combination with its own release time, and enables any party to verify the released result without trusted setup or costly asymmetric-key proof systems. Technically, our construction maps independently generated RSA-based puzzles into a common finite field, uses oblivious linear evaluation to refresh blinding factors during evaluation, and embeds a hidden verification structure by encoding messages as polynomials with committed secret roots. We formalize verifiable homomorphic linear-combination TLPs, prove privacy and solution validity in this model, and capture the setting in which the evaluation result may be released before the underlying client puzzles are opened. Our prototype implementation shows that verifying an evaluated result takes less than 3\,ms.



\end{abstract}



\maketitle

\section{Introduction}\label{sec::intro}

In many settings, parties need to lock sensitive values now so that they become available only after a designated delay, without retaining those values locally in the meantime~\cite{TimothyMay1993,Rivest:1996:TPT:888615,xin2024check}. This need arises naturally in distributed sensing, drone, and cyber-physical deployments operating in remote, hostile, or security-sensitive environments~\cite{YaacoubSNKCM20,RawatSCB14,MitchellC13}. In such settings, a sensor, drone, site, or local operator may wish to protect a measurement, incident count, or log-derived statistic immediately, deleting the plaintext and retaining only the secret state needed for later authorization, rather than leaving the plaintext value itself on an exposed device that may later be inspected, seized, captured, or otherwise compromised. 
Crucially, delayed recovery should not depend on the owners later returning to enable that release. This requirement is especially important in hostile or exposed deployments, where devices or operators may not remain reliably available at the intended release time.
At the same time, an authorized aggregate, such as a total count, average measurement, weighted risk score, or threshold outcome, may need to be released and publicly verified before the underlying reports are disclosed.

\smallskip\noindent
This combination of requirements arises in concrete settings. In cross-organizational incident reporting, organizations time-lock incident counts or risk metrics as they occur; after the reporting window, they authorize an aggregate risk score with a short delay, enabling early sector-wide response, while individual reports, which could expose a victim or an ongoing investigation, remain locked far longer. Similar patterns arise in timed surveys, where aggregate statistics can be published early while individual responses remain protected. 
As these examples suggest, parties benefit from heterogeneous delays for several reasons. First, disclosure constraints are rarely uniform: organizations face different legal, contractual, or forensic embargo periods, so each fixes its own delay. Second, clients generate puzzles at different times, so even a common disclosure date requires distinct delays. Third, the aggregate is far less sensitive than its constituents, so it warrants a much shorter delay.



Homomorphic time-lock puzzles~\cite{MalavoltaT19} provide a natural starting point for this setting. They allow parties to encode sensitive inputs into puzzles, publish them, and then stop relying on long-term local retention of those values. Once the puzzles are published, a server can derive a new puzzle whose solution represents a function of the underlying hidden inputs~\cite{BrakerskiDGM19,liu2022towards,ThyagarajanCLM21}.  Linearly homomorphic variants are particularly appealing here,
since many desired outputs, including sums, counts, averages, and
threshold outcomes, reduce to linear combinations. 
Yet existing homomorphic TLPs leave a critical part of this setting unresolved. Prior homomorphic TLP schemes do not provide an efficient mechanism for publicly verifying that the released evaluation result was computed correctly~\cite{MalavoltaT19,BrakerskiDGM19,liu2022towards,ThyagarajanCLM21}. A server may output a puzzle that claims to encode the desired aggregate, but the parties have no effective way to check whether the intended inputs, subset, and coefficients were used honestly. As a result, the parties would be forced to trust it where trust should be least acceptable. 

One possible remedy is to postpone verification until the underlying puzzles are solved and then recompute the aggregate from the revealed plaintexts. However, that does not address the actual requirement. The aggregate may need to be released and publicly verified long before the underlying inputs are opened; for example, an evaluation puzzle may be intended for release within months, whereas the individual client puzzles may remain locked for years. Waiting for all underlying puzzles to be solved would therefore defeat the purpose of early aggregate release. 
Another alternative is homomorphic timed commitments~\cite{ChvojkaJ23}, in which evaluation is a public deterministic operation that anyone can recompute, and forced opening is efficiently verifiable. However, all commitments there share a single delay fixed by a trusted setup, and any party can force open any commitment in that same time; hence, this approach cannot release a verified aggregate under a shorter delay while the underlying inputs remain protected under longer delays.

A natural question is whether existing homomorphic TLPs can be extended in a lightweight way to support efficient public verification. This turns out to be nontrivial. On the one hand, verification must certify correctness before the underlying puzzles are solved, since the aggregate may be intended for release much earlier. On the other hand, this must be achieved efficiently and in a publicly verifiable manner, without relying on costly asymmetric-key proof systems (e.g., zero-knowledge proofs) or stronger assumptions such as trusted setup, both of which would undermine practicality.

\vspace{-3mm}

\subsection{Our Contributions}

We introduce \tf, the first homomorphic TLP
scheme that supports \emph{efficient public verification} of both (i) a homomorphic linear combination of puzzles from multiple clients and (ii) a single client's puzzle solution. Our main contributions are as follows.

%
%

\noindent$\bullet\ $ \textbf{\tf: the first verifiable homomorphic linear-combination TLP.}
We construct \tf, a scheme that lets clients generate and publish their puzzles independently and later authorize a desired linear combination over them, without relying on trusted setup or costly asymmetric-key proof systems. \tf considers a strong malicious server that may gain access to a subset of clients' secret parameters, and still ensures the privacy of honest clients and the validity of the released result.

\noindent$\bullet\ $ \textbf{New techniques.}
The main technical contribution of \tf is a new way to embed
verifiability into homomorphic TLP evaluation using tools that, to our knowledge, have not been combined in prior TLP schemes. At a high level, we encode each message as a polynomial, protect it using a verifiable masked polynomial, and use oblivious linear evaluation to switch blinding factors so that homomorphic evaluation remains secure even across repeated computations. Our design is efficient, since the homomorphic operations are carried out over a relatively small finite field, e.g., 128-bit.

\noindent$\bullet\ $ \textbf{New verifiable homomorphic TLP model.}
We formalize verifiable homomorphic linear combination TLPs and analyze the security of \tf within this model. Our definition also captures a setting in which the evaluation result may be released before the underlying client puzzle solutions are revealed. To the best of our knowledge, this earlier-release setting is not explicitly captured in prior homomorphic TLP privacy definitions.

\noindent$\bullet\ $ \textbf{Prototype implementation and concrete evaluation.} 
We implemented \tf and report its concrete cost. In our
experiments, verifying an evaluated result takes less than 3\,ms.

\subsection{Technical Overview}

Designing a verifiable homomorphic TLP in our setting raises three
main challenges. 

First, homomorphic evaluation requires all puzzles to live over a common algebraic domain, yet we do not want to assume a trusted
setup or require all clients to share the same RSA modulus. To
address this, each client generates its own RSA modulus, as in the
original RSA-based TLP setting, but encodes its puzzle contents in a shared finite field. Concretely, each client derives pseudorandom masking values from its time-lock key and uses them to represent its message as a field element. This allows clients to prepare puzzles independently, without a trusted setup, and still enables later homomorphic operations across those puzzles.

\begin{figure}[!tb]
    \centering
    \includegraphics[width=84mm]{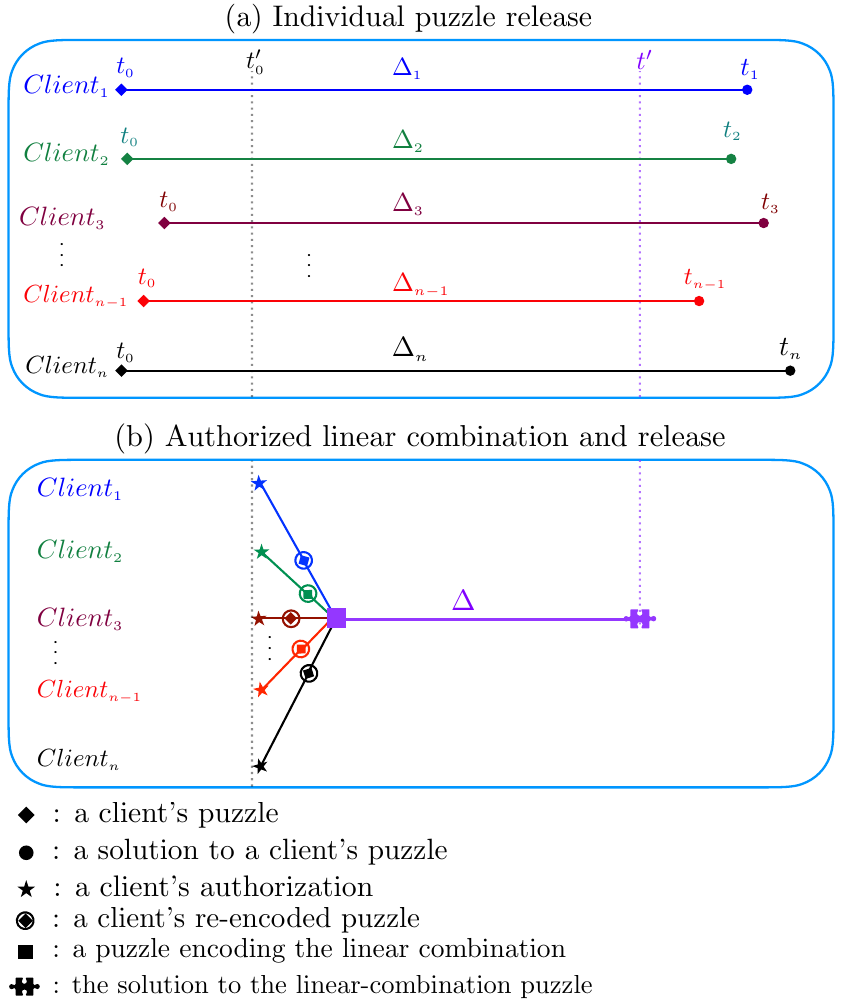}
        \vspace{-2.5mm}
    \caption{
    Overview of \tf. Clients first publish individual puzzles and may later authorize a linear combination with its own release time $t'$.}
    \label{fig:parties-interactions-in-TF}
    \vspace{-4mm}
\end{figure}

Second, a homomorphic linear combination cannot be obtained by
simply summing the published puzzles, because each client protects
its puzzle using independent blinding factors. A direct combination
would therefore not decode to the desired result. Our solution is to
refresh these blinding factors during evaluation. We do this using
oblivious linear evaluation ($\ole^{\st +}$), which lets the server obtain
properly re-encoded values without learning the underlying
plaintexts. This re-encoding step also helps preserve security across
repeated evaluations.

Third, public verification is challenging because the verifier must
check the released evaluation result before the underlying puzzles are opened. To achieve this without relying on heavyweight
asymmetric-key proof systems, we encode messages as polynomials
and embed a hidden verification structure into the evaluated result. Specifically, selected clients introduce secret roots into the polynomial encoding, commit to those roots, and ensure that a correctly generated evaluated puzzle preserves them. After the server solves the evaluated puzzle, it releases the  result along with openings for the committed roots. A verifier can reconstruct the corresponding polynomial, validate the openings, and check that the published output is consistent with the intended linear combination.

At a high level, \tf works as follows. Each client first
publishes a puzzle encoding its secret message. Later, a  set of clients can ask the server to derive a new puzzle whose solution is a linear combination of their still-hidden messages, with its own release time. After the designated delay, the server publishes the solution and proof material for an evaluated puzzle or a single-client puzzle, and any party can verify the released output using only public information. Figure~\ref{fig:parties-interactions-in-TF} outlines the workflow of \tf.



\begin{table*}[!t]
\begin{center}
\caption{Comparing features and costs of different (partially) homomorphic TLPs. Setup includes all pre-evaluation costs, including global setup and puzzle generation. Entries omit the outer $O(.)$ for brevity. $T$ is the time parameter, 
$\lambda$ is the security parameter, 
$n$ is total number of puzzles, \tl is total number of leaders, and $|\mathcal{F}|$ is the size of the evaluated function/circuit.}
\label{feature-of-our-schemes} 
\vspace{-2mm}

\renewcommand{\arraystretch}{1.1}
\scalebox{.94}{
\begin{tabular}{cc|c|c|c|c|c|c|c|cc} 
   \hline
   &
   &Client
   &  No&\multicolumn{2}{c|}{Verification}&  Separate Eval.&\multicolumn{3}{c}{Asymptotic Complexity}
   \\
   \cline{5-6}
   \cline{8-10}
 \multirow{-2}{*}{Scheme}
 &
 &\multirow{-1}{*}{in Eval.}
 &\multirow{-1}{*}{   Trusted-Setup}&{  Client's Solution}&{Eval.  Result}&\multirow{-1}{*}{Release Time}&Setup&Eval.& Eval. Verification
 \\
\hline
 {\rotatebox[origin=c]{0}{  \textbf{Ours}}}
 &
 &\cellcolor{white!20}{{  \checkmark}}
 &\cellcolor{white!20}{{  \checkmark}}
 &\cellcolor{white!20}{{  \checkmark}}
&\cellcolor{white!20}{{  \checkmark}}
&\cellcolor{white!20}{{  \checkmark}}
&$\log T$
&$poly(\log T, \lambda, n, \tl)$
&$\tl^{\st 2}$
 \\      


\hline
{\rotatebox[origin=c]{0}{   \cite{BrakerskiDGM19}}}
&
&\cellcolor{white!20}{\rotatebox[origin=c]{0}{  $\times$}}
&\cellcolor{white!20}{{  $\times$}}
&\cellcolor{white!20}{{  $\times$}}
&\cellcolor{white!20}{{  $\times$}}
&\cellcolor{white!20}{{  $\times$}}
&$\log T$
&$poly( \lambda, n, |\mathcal{F}|)$
&--
\\

   
\hline
{\rotatebox[origin=c]{0}{   \cite{liu2022towards}}}
&
&\cellcolor{white!20}{\rotatebox[origin=c]{0}{  $\times$}}
&\cellcolor{white!20}{{  $\times$}}
&\cellcolor{white!20}{{  $\checkmark$}}
&\cellcolor{white!20}{{  $\times$}} 
&\cellcolor{white!20}{{  $\times$}}
&$\log T$
&$poly(\lambda, n)$
&--
\\
\hline

{\rotatebox[origin=c]{0}{   \cite{MalavoltaT19}}}
&
&\cellcolor{white!50}{\rotatebox[origin=c]{0}{  $\times$}}
&\cellcolor{white!50}{{  $\times$}}
&\cellcolor{white!50}{{  $\times$}}
&\cellcolor{white!50}{{  $\times$}}
&\cellcolor{white!50}{{  $\times$}}
&$\log T$
&$poly( \lambda, n)$
&--
\\

\hline
{\rotatebox[origin=c]{0}{   \cite{ThyagarajanCLM21}}}
&
&\cellcolor{white!50}{\rotatebox[origin=c]{0}{  $\times$}}
&\cellcolor{white!50}{{  $\checkmark$}}
&\cellcolor{white!50}{{  $\times$}}
&\cellcolor{white!50}{{  $\times$}}&\cellcolor{white!50}{{  $\times$}}
&$T$&$poly(\lambda, n)$
&--
 \\


 \hline
\end{tabular}
}
\end{center}
\end{table*} 
\section{Related Work}\label{Related-Work}

\noindent\textbf{Time-Lock Puzzles.} 
Timothy May~\cite{TimothyMay1993} was the first to propose sending information into the future via time-lock puzzles. May's construction relied on a trusted agent to release the secret. Later, Rivest \et.~\cite{Rivest:1996:TPT:888615} introduced an RSA-based TLP
that requires no trusted agent, relies on sequential modular squaring, and resists parallel computation. A basic property of such schemes is that puzzle generation is faster than solving. 
Since then, many variants have been developed, including multi-instance TLPs~\cite{AbadiK21}, 
post-quantum TLPs~\cite{LaiM24,crypto-2024-34246}, and homomorphic TLPs~\cite{MalavoltaT19,BrakerskiDGM19,liu2022towards}. 
Malavolta and Thyagarajan~\cite{MalavoltaT19} introduced homomorphic TLPs and gave candidate constructions for different homomorphism classes; their fully homomorphic TLP relied on indistinguishability obfuscation (iO). Brakerski \et.~\cite{BrakerskiDGM19} later obtained a fully homomorphic TLP from standard assumptions.

To improve efficiency, partially homomorphic TLPs have
been proposed, supporting operations such as homomorphic linear combinations or  multiplication~\cite{MalavoltaT19,liu2022towards}. Unlike the partially
homomorphic TLP in~\cite{MalavoltaT19}, the construction in~\cite{liu2022towards}
additionally allows a verifier to (1)~check correct puzzle generation and (2)~verify a server's solution for a single client's puzzle (but not homomorphic computations), using a public-key proof~\cite{Wesolowski19}. 
Thyagarajan \et.~\cite{ThyagarajanCLM21} showed that linearly homomorphic TLPs can also be instantiated from class groups of imaginary quadratic order. However, this is achieved at the cost of a one-time global setup phase whose running time is proportional to the delay parameter $T$. Thus, if $T$ is calibrated to correspond to roughly two years of sequential work for puzzle solving, then generating the global public parameters would also require roughly the same order of offline sequential work.

All these homomorphic TLPs (except \cite{ThyagarajanCLM21}) require a trusted setup, 
and none support verifying the correctness of homomorphic computations.  These schemes treat evaluation as a public transformation on published puzzles. Once the puzzles are published, any party can run supported functions on them without further involvement from the puzzle owners. They have not been designed to capture a setting in which clients can later authorize computation over their puzzles and specify a distinct release time for the computed result. 
Table~\ref{feature-of-our-schemes} outlines the features and costs of these homomorphic TLPs.

A related but distinct line of work studies batchable TLPs~\cite{SrinivasanLMNPT23,dujmovic2023time}. Unlike homomorphic TLPs, these interesting schemes are not designed to release a function value computed over hidden puzzle contents. Instead, they combine many puzzles into a single puzzle whose solution reveals all underlying messages at once. 

\noindent\textbf{Timed Commitments.}  
Timed commitments~\cite{BonehN00} augment commitments with forced opening, and modern constructions add non-malleability and efficient public verification~\cite{ChvojkaJ23,ThyagarajanCLM21,KatzLX20,AbusalahA24}. Notably, Chvojka and Jager~\cite{ChvojkaJ23} construct non-malleable timed commitments with a linear (or multiplicative) homomorphism: evaluation is a deterministic public operation, so any party can verify an aggregated commitment by recomputing it, and a forced-opening proof certifies its opening without the $T$-step computation. This setting, however, differs from ours. All commitments live under a common reference string generated by a trusted setup for a single hardness parameter~$T$, so clients cannot choose individual delays; and since any party can force open any individual commitment in the same time~$T$, an aggregate cannot be released under a shorter delay while the underlying values remain locked under longer delays. In contrast, \tf requires no trusted setup, supports client-chosen delays, and releases an authorized evaluation long before any input is released; the price is that evaluation involves client secrets and is not publicly recomputable, which is why an explicit verification mechanism is needed. These schemes also offer non-malleability, which is essential for fair MPC and orthogonal to our goal.

%


\section{Preliminaries}\label{sec::prelminaries}

\subsection{Notations and Informal Threat Model}\label{notations}

We define $\Delta_{\st \prt}$ as the period within which client $\prtt_{\st \prt}$ wants its puzzle's solution $m_{\st \prt}$ to remain secret. 
Let \mxs denote the upper bound on the number of sequential computation steps that a solver with the highest level of computation resources can perform per second.
In the case of classical RSA-based time-lock puzzles and our proposed protocol, we specialize this to \mxsqr, representing the maximum number of modular squarings a solver can execute per second. 

We define $U$ as the universe of a solution $m_{\st\prt}$. 
%
%
We denote by $\lambda\in \mathbb{N}$ the security parameter. We
use the polynomial $poly(\lambda)$ to state that the parameter is a polynomial
function of $\lambda$. 

We define a hash function
$g: \{0,1\}^{\st *} \rightarrow \{0,1\}^{\st poly(\lambda)}$. We denote a
null value or set by $\bot$. By $||v||$ we mean the bit-size of $v$ and by
$||\vec{v}||$ we mean the total bit-size of elements of $\vec{v}$. We denote by
$\prm$ a large prime number, where $\log_{\st 2}(\prm)$ is the security
parameter, e.g., $\log_{\st 2}(\prm) \geq 128$. For correctness, we assume that \prm is chosen sufficiently large so that no relevant intermediate value wraps around for any input in the range. 

In our scheme, the server's public parameters are
$pk_{\st s} \asn (\prm, X)$, where $X \asn \{x_{\st 1},\ldots,x_{\st \pt} \}
\subseteq \F_{\st \prm}$. We define $\valSetup(pk_{\st s})$ as a deterministic polynomial-time predicate that outputs $1$ if: (1) $\prm$ is prime and has the prescribed bit-length; (2) the elements of $X$ are pairwise distinct; (3) each $x_{\st i} \in X$ is nonzero and $x_{\st i} \notin U$. Otherwise, it outputs $0$.

To ensure generality in the definition of our verification algorithms, we adopt
notations from zero-knowledge proof systems~\cite{BlumSMP91,FeigeLS90}. Let
$R_{\st \cmd}$ be an efficient binary relation consisting of pairs
$(stm_{\st \cmd}, wit_{\st \cmd})$, where $stm_{\st \cmd}$ is a statement and
$wit_{\st \cmd}$ is a witness. Let $\mathcal{L}_{\st \cmd}$ be the language (in
$\mathcal{NP}$) associated with $R_{\st \cmd}$. In this paper, two main types
of verification occur: (1)~verification of a single client's puzzle solution,
$\cmd=\scp$, and (2)~verification of a linear combination, $\cmd=\ep$. 
We assume that all point-to-point communications between parties use authenticated and confidential channels.


We define $C\asn\set$ as a set of clients. The scheme
designates (at random) a subset of clients as leaders. Let \idx be this subset,
containing $\tl$ leaders. 
We consider a strong malicious server and semi-honest clients. This asymmetry is deliberate and mirrors outsourced computation: the participating clients are known entities, such as organizations or managed devices, that themselves initiate and authorize the evaluation, and thus have little incentive to deviate from a computation they requested, whereas the evaluation is delegated to an untrusted server, where misbehavior is most likely and most consequential. Although clients follow the protocol, the model still captures client-state exposure and passive collusion:  the adversary controls the server and can obtain the secret keys and local parameters of a subset $\corupt$ of clients, including some leaders.


We define a
threshold $t$ and require the number of corrupted leaders to be at most 
$t$, i.e., $|\corupt \cap \idx| \leq t$.  We define $\cor\asn\tl+2$. 


\subsection{Pseudorandom Function}\label{sec::prf}

A pseudorandom function is a deterministic keyed function whose
output is computationally indistinguishable from that of a truly
random function~\cite{DBLP:books/crc/KatzLindell2007}. In this
paper, we use: 
$
\mathtt{PRF}:\{0,1\}^{\st *}\times\{0,1\}^{\st poly(\lambda)}
\rightarrow \F_{\st \prm}.
$  
We write $\mathtt{PRF}(x,k)$ for the evaluation of $\mathtt{PRF}$ on input $x$ and secret seed $k$. We use $\mathtt{PRF}$ to derive pseudorandom field elements, including the values used to blind encoded messages and related auxiliary masks. Whenever a value derived from \prf is used as a multiplicative mask, it is taken to be an element of $\F_{\st \prm}^{\st *}$, and hence is
nonzero.

\subsection{Oblivious Linear  Evaluation}\label{sec::OLE-plus}

Oblivious Linear Evaluation (\ole) is a two-party protocol that
involves a sender and receiver. In \ole, the sender has two inputs
$a, b\in \F_{\st \prm}$ and the receiver has a single input,
$c \in \F_{\st \prm}$. The protocol allows the receiver to learn only
$s \asn a\cdot c + b \in \F_{\st \prm}$, while the sender learns nothing. Ghosh \et.~\cite{GhoshNN17} proposed an efficient \ole that has $O(1)$
overhead. Later, in~\cite{GhoshN19} an enhanced \ole, called $\ole^{\st +}$, was proposed. The latter ensures that the receiver cannot learn anything about the sender's
inputs, even if it sets its input to $0$. $\ole^{\st +}$ is also accompanied by
an efficient symmetric-key-based verification mechanism that enables a party to
detect its counterpart's misbehavior during the protocol's execution. We use
$\ole^{\st +}$ to securely switch the blinding factors of secret messages
encoded in puzzles. We refer readers to Appendix~\ref{apndx:F-OLE-plus} for
the construction of $\ole^{\st +}$.

\subsection{Polynomial Representation of a Message}%
\label{sec:Polynomial-Representation-of-Message}

Encoding a message $m$ as a polynomial ${\pi}(x)$ allows us to impose a
certain structure on the message. Polynomial representation has been used in
various contexts, such as secret sharing~\cite{Shamir79}, private set intersection~\cite{DBLP:conf/crypto/KissnerS05}, or error-correcting codes~\cite{reed1960polynomial-}. 
There are two common approaches to encode $m$
in ${\pi}(x)$:

\begin{enumerate}[leftmargin=4.9mm]
  \item Set $m$ as the constant term of ${\pi}(x)$, e.g.,
    ${\pi}(x)\asn m+\sum_{\st j=1}^{\st n}x^{\st j}\cdot a_{\st j}$.
  \item Set $m$ as the root of ${\pi}(x)$, e.g.,
    ${\pi}(x)\asn(x-m)\cdot \tau(x)$.
\end{enumerate}

We will use both approaches. The former lets us perform a linear combination of
the constant terms of different polynomials. We utilize the latter to insert a
secret random root, acting as a ``trap'', into the polynomials encoding the
messages. The resulting polynomial representing the linear combination contains
this specific root, facilitating verification of correctness.

\noindent\textbf{Point-Value Form.} A polynomial $\pi(x)$ of degree $d$ can
be represented as a set of $\pt$ ($\pt>d$) point-value pairs $\{(x_{\st 1},
\pi_{\st 1}),\ldots,(x_{\st \pt},\pi_{\st \pt})\}$ where all $x_{\st i}$ are
distinct non-zero points and $\pi_{\st i}=\pi(x_{\st i})$ for all $i$. A
polynomial in this form can be converted into coefficient form via polynomial
interpolation, e.g., Lagrange interpolation~\cite{aho19}.

\noindent\textbf{Arithmetic of Polynomials in Point-Value Form.} Arithmetic of
polynomials in point-value representation is done by adding or multiplying the
corresponding $y$-coordinates. Let $a$ be a scalar and
$\{(x_{\st 1},\pi_{\st 1}),\ldots,(x_{\st \pt},\pi_{\st \pt})\}$ be $(x,y)$-coordinates
of a polynomial ${\pi}(x)$. The polynomial ${\theta}$ interpolated from
$\{(x_{\st 1}, a\cdot\pi_{\st 1}),\ldots,(x_{\st \pt},a\cdot\pi_{\st \pt})\}$ is
the product of $a$ and ${\pi}(x)$, i.e., ${\theta}(x)=a\cdot {\pi}(x)$.

\subsection{Verifiable Masked Polynomial}%
\label{Unforgeable-Encrypted-Polynomial}

An interesting property of masked polynomials was stated
in~\cite{AbadiTD16}. Let ${\pi}(x)$ be a polynomial with a random secret
root $\beta$, represented in point-value form with its $y$-coordinates
masked. Given the $x$-coordinates and masked $y$-coordinates, the
probability that an adversary can modify a subset of the masked $y$-coordinates such that, after unmasking and interpolating, the resulting
polynomial still has root $\beta$ is negligible. Below, we formally state it.

\begin{theorem}\label{theorem::Unforgeable-Encrypted-Polynomial}
Let ${\pi}(x)$ be a degree-$d$ arbitrary polynomial. Define $\zeta(x)$ as a degree-$1$ polynomial with a   root $\roott$ sampled uniformly from $\F_{\st \prm}$, i.e.,  
$\zeta(x)\asn (x-\roott)$. 
Define $\theta(x) \asn \pi(x)\cdot \zeta(x)$ over $\F_\prm[x]$. Let   
$ \big\{(x_{\st  1},y_{\st  1}),\ldots, (x_{\st\pt}, y_{\pt})\big\}$ be the point-value representation of $ \theta(x)$, where $\pt> d+1$ and   $x_{\st 1} ,\ldots, x_{\st\pt}$  are distinct non-zero values of $\ \F_{\st \prm}$. 
 Define  the encoding of each $y$-coordinate $y_{\st  i}$ of $\theta(x)$ as: 
$\enco_{\st  i} \asn w_{\st  i}\cdot (y_{\st  i}+z_{\st  i}) $ 
where  $w_{ \st i}$ and $z_{\st  i}$ are chosen uniformly at random from $\F^{\st *}_{\st  \prm}$ and $\F_{\st  \prm}$ respectively. 
Let $\vec{\enco}' \asn [{\enco}'_{\st  1}, \ldots, {\enco}'_{\st\pt}]$, where $\exists {\enco}'_{\st i}\in \vec{\enco}'$ such that ${\enco}'_{\st  i}\neq {\enco}_{\st  i}$. Define  $\psi(x)$ as a polynomial interpolated from decoded $y$-coordinates:  
$\Big\{\big(x_{\st  1}, (w^{\st  -1}_{\st  1}\cdot {\enco}'_{\st  1})-z_{\st  1}\big), \ldots, \big(x_{\st\pt}, (w^{\st  -1}_{\st\pt}\cdot {\enco}'_{\st\pt})-z_{\st\pt}\big)\Big\}.$ 
Then, the probability that $\psi(x)$ has the root \roott  is at most $\frac{\pt-1}{\prm}$, i.e., 
$
\Pr\Big[ \psi(\roott) =0\Big]\leq \frac{\pt-1}{\prm}. 
$
\end{theorem}



Appendix~\ref{sec::proof-of-unforgeable-poly} proves
Theorem~\ref{theorem::Unforgeable-Encrypted-Polynomial}.

\subsection{Commitment Scheme}\label{subsec:commit-prelim}

A commitment scheme involves a \emph{sender} and a \emph{receiver}, with two
phases: \emph{commit} and \emph{open}. In the \emph{commit} phase, the sender
commits to a message $x$ as $com \asn \mathtt{Com}(x,r)$, involving a secret value
$r$. In the \emph{open} phase, the sender sends the opening
$\ddot{x}:=(x,r)$ to the receiver, which verifies its correctness:
$\mathtt{Ver}(com,\ddot{x})\stackrel{\scriptscriptstyle ?}=1$ and accepts if
the output is $1$. A commitment scheme satisfies: (a)~\textit{hiding}: it is
infeasible for an adversary to learn any information about $x$ until $com$ is
opened, and (b)~\textit{binding}: it is infeasible for a malicious sender to
open $com$ to a value different from the one used in the commit phase.
Appendix~\ref{subsec:commit} presents further details.

\subsection{Time-Lock Puzzle}\label{sec::Time-lock-Encryption}
In this section, we restate the definition of a time-lock puzzle (TLP).  
\begin{definition}\label{Def::Time-lock-Puzzle}
A TLP consists of $(\mathsf{Setup_{\st TLP}},\mathsf{GenPuzzle_{\st TLP}},
\mathsf{Solve_{\st TLP}})$. It meets completeness and efficiency properties.
TLP involves a client \prtt and a server $\srv$.

  \noindent$-$ $\mathsf{Setup_{\st TLP}}(1^{\st\lambda},\Delta, \mxs)\rightarrow
    (pk,sk)$. A probabilistic algorithm, run by \prtt. It takes as input a
    security parameter $1^{\st \lambda}$, a time parameter $\Delta$ specifying
    how long a message must remain hidden, and a time parameter $\mxs$ which
    is the maximum number of computational steps a solver can perform per second.
    It outputs a pair $(pk,sk)$.

  \noindent$-$ $\mathsf{GenPuzzle_{\st TLP}}(m, pk, sk)\rightarrow {o}$. A
    probabilistic algorithm, run by \prtt. It takes as input a solution $m$
    and $(pk,sk)$. It outputs a puzzle $o$.

  \noindent$-$ $\mathsf{Solve_{\st TLP}}(pk, {o})\rightarrow s$. A deterministic
    algorithm, run by $\srv$. It takes as input $pk$ and ${o}$. It outputs a
    solution $s$.

A TLP scheme offers three core properties, completeness, Efficiency, and security, defined below.  

\noindent$\bullet$ \textit{Completeness:}  For honest \prtt and $\srv$, it holds that
  $\mathsf{Solve_{\st TLP}}(pk, o)=m$.

\noindent$\bullet$ \textit{Efficiency:} The runtime of $\mathsf{Solve_{\st TLP}}$ is upper bounded by $poly(\Delta,\lambda)$.
\end{definition}

The security of a TLP requires that the puzzle's solution remains private from all
adversaries running in parallel within the time period $\Delta$.
\begin{definition}
A TLP is secure if for all $\lambda$ and $\Delta$, all PPT adversaries
$\mathcal{A}:=(\mathcal{A}_{\st 1},\mathcal{A}_{\st 2})$ where
$\mathcal{A}_{\st 1}$ runs in total time $O(poly(\Delta,\lambda))$ and
$\mathcal{A}_{\st 2}$ runs in time $\delta(\Delta)<\Delta$ using at most
$\xi(\Delta)$ parallel processors, there is a negligible function $\mu$, such
that:
$$\Pr\left[
  \begin{array}{l}
\mathsf{Setup_{\st TLP}}(1^{\st\lambda},\Delta, \mxs)\rightarrow (pk, sk)\\
\mathcal{A}_{\st 1}(1^{\st\lambda},pk, \Delta)\rightarrow (m_{\st 0},m_{\st 1},\text{state})\\
b\stackrel{\st\$}\leftarrow \{0,1\}\\
\mathsf{GenPuzzle_{\st TLP}}(m_{\st b}, pk, sk)\rightarrow {o}\\
\hline
\mathcal{A}_{\st 2}(pk, {o},\text{state}) \rightarrow b \\
\end{array}
\right]\leq \frac{1}{2}+\mu(\lambda)$$
\end{definition}

We refer readers to Appendix~\ref{sec::RSA-based-TLP} for the description of
the original RSA-based TLP. The definitions
of sequential and iterated sequential functions in the presence of an adversary
possessing a polynomial number of processors are restated in Appendix~\ref{sec::equential-squering}.




\section{Model}\label{sec::definition}





In general, the basic functionality $\func$, that any $n$-party Private Linear Combination (PLC) computes, takes as input a pair of coefficient $q_{\st j}$ and plaintext value $m_{\st j}$ from the $j$-th party (for every $j, 1\leq j \leq n$), and returns their linear combination $\sum\limits^{\st n}_{\st j=1}q_{\st j}\cdot m_{\st j}$ to each party. More formally, $\func$ is defined as:
 \vspace{-1mm}
 \begin{equation}\label{equ::PLC}
\func \big((q_{\st 1}, m_{\st 1}),\ldots,(q_{\st n}, m_{\st n}) \big)\rightarrow (q_{\st 1}\cdot m_{\st 1}+\ldots+q_{\st n}\cdot m_{\st n})
\end{equation}

$\func$ implies that corrupt colluding parties can always deduce the linear combination of non-colluding parties' inputs from the output of $\func$. This point holds for any scheme that realizes this functionality (including the ones in \cite{MalavoltaT19,BrakerskiDGM19} and ours), regardless of the primitives used.  Next, we present a definition of Verifiable Homomorphic Linear Combination TLP ($\vhtlp$), by initially presenting the syntax and then providing the security and correctness definitions.


\subsection{Syntax}


\begin{definition}[Syntax] A Verifiable Homomorphic Linear Combination TLP (\vhtlp) scheme consists of six algorithms: $(\ssetup, $ $\csetup,$ $ \pgen, \eval, \solv,$ $ \ver)$, defined below.


\noindent$\bullet$  $\ssetup(1^{\st \lambda}, \tl)\rightarrow pk_{\st \srv}$. An algorithm, run by the server \srv. It takes as input security parameters $1^{\st \lambda}$ and $\tl$. It returns public parameters $pk_{\st \srv}$. 
 Server \srv publishes $pk_{\st \srv}$. 

\noindent$\bullet$  $\csetup(1^{\st \lambda})\rightarrow \kp_{\st \prt}$.  An algorithm run by a client $\prtt_{\st\prt}$. It takes a security parameter $1^{\st \lambda}$ as input. It returns a pair $\kp_{\st \prt}\asn(sk_{\st \prt}, pk_{\st \prt})$ of secret key $sk_{\st \prt}$ and  public key $pk_{\st \prt}$. Client $\prtt_{\st\prt}$ publishes $pk_{\st \prt}$.

\noindent$\bullet$  $\pgen(m_{\st \prt}, \kp_{\st \prt}, pk_{\st \srv}, \Delta_{\st \prt}, \mxs)\rightarrow(\vec{o}_{\st \prt}, prm_{\st \prt})$.  An algorithm,  run by $\prtt_{\st\prt}$. It takes as input plaintext message $m_{\st \prt}$, key pair $\kp_{\st \prt}$, server's public parameters $pk_{\st \srv}$, time parameter $\Delta_{\st \prt}$ determining the period during which $m_{\st \prt}$ should remain private, and the maximum number $\mxs$ of computational steps per second that the strongest solver can perform. 
%
%
It returns a puzzle vector $\vec{o}_{\st \prt}$ (representing a single puzzle) along with a pair  $prm_{\st \prt}\asn (sp_{\st \prt}, pp_{\st \prt})$ of secret parameter $sp_{\st \prt}$ and public  parameter $pp_{\st \prt}$ of the puzzle, which may include the index of $\prtt_{\st\prt}$.  Client $\prtt_{\st\prt}$ publishes $(\vec{o}_{\st \prt}, pp_{\st \prt})$.

\noindent$\bullet$  $\eval(\langle \srv(\vec{o}, \Delta,  \vec{pp}, \vec{pk}, pk_{\st \srv}), c_{\st 1}(\Delta,  \kp_{\st {\st 1}}, prm_{\st {\st 1}},  q_{\st 1}, $ $ pk_{\st \srv}), $ $\ldots,$ $ c_{\st n}(\Delta, $ $\kp_{\st {\st n}}, $\\ $prm_{\st {\st n}}, q_{\st n},$ $pk_{\st \srv}) \rangle)\rightarrow(\vec{g}, \vec{pp}^{\st(\text{Evl})})$. An (interactive) algorithm, run by  \srv (and each client in $\set$). When no interaction between \srv and the clients is required, the clients' inputs will be null $\bot$. Server \srv takes as input vector $\vec{o}$ of $n$ clients' puzzles, time parameter $\Delta$ within which the evaluation result should remain private (where $\Delta<min(\Delta_{\st 1},\ldots, \Delta_{\st n})$),    $\vec{pp} \asn [pp_{\st 1},\ldots, pp_{\st n}]$,  $\vec{pk}\asn [pk_{\st 1},\ldots,$  $pk_{\st n}]$, and  $pk_{\st \srv}$. Each $\prtt_{\st\prt}$ takes as input $\Delta,   \kp_{\st \prt}, prm_{\st \prt}$, coefficient $q_{\st \prt}$, and $pk_{\st \srv}$. It returns a vector of public parameters $\vec{pp}^{\st (\text{Evl})}$ and a puzzle vector  $\vec{g}$, representing a single puzzle.  Server \srv publishes $\vec{g}$ and the clients publish $\vec{pp}^{\st (\text{Evl})}$. 

\noindent$\bullet$  $\solv(\vec{o}_{\st \prt}, pp_{\st \prt}, \vec{g}, \vec{pp}^{\st (\text{Evl})}, \vec{pk}, pk_{\st \srv}, \cmd)\rightarrow(m, \zeta)$. An algorithm run by \srv. It takes as input a client $\prtt_{\st\prt}$'s puzzle vector $\vec{o}_{\st \prt}$, the puzzle's public parameters $pp_{\st \prt}$, a vector $ \vec{g}$ representing the puzzle that encodes evaluation of all clients' puzzles, a vector of public parameter $\vec{pp}^{\st (\text{Evl})}$, $\vec{pk}$, $pk_{\st \srv}$, and a command \cmd, where $cmd \in \{\text{\scp, \ep}\}$. When $\cmd=\scp$, it solves puzzle $\vec{o}_{\st \prt}$ (which is an output of $\pgen $), this yields a solution  $m$. In this case, input $\vec{g}$ can be null $\bot$. When $\cmd=\ep$, it solves puzzle $\vec{g}$ (which is an output of $\eval $), this yields a solution $m$.  In this case, input $\vec{o}_{\st \prt}$ can be $\bot$. Depending on the value of \cmd, it generates a proof $\zeta$ for the corresponding statement in $\mathcal{L}_{\st \cmd}$. It outputs a solution $m$ and a proof $\zeta$. Server \srv publishes $(m, \zeta)$.

\noindent$\bullet$  $\ver(m, \zeta, \vec{o}_{\st \prt}, pp_{\st \prt}, \vec{g}, \vec{pp}^{\st (\text{Evl})}, pk_{\st \srv}, \cmd)\rightarrow \ddot{v}\in\{0,1\}$. An algorithm run by a verifier. It takes as input a plaintext solution $m$, proof $\zeta$, puzzle $\vec{o}_{\st \prt}$ of a single client $\prtt_{\st \prt}$, 
public parameters  $pp_{\st \prt}$ of  $\vec{o}_{\st \prt}$,  a puzzle $\vec{g}$ for a linear combination of puzzles, public parameters $\vec{pp}^{\st (\text{Evl})}$ of $\vec{g}$, server's public key $pk_{\st \srv}$, and command \cmd that determines whether the verification corresponds to $\prtt_{\st \prt}$'s single puzzle or linear combination of puzzles. It returns $1$ if it accepts the proof. It returns $0$ otherwise.  
%

\end{definition}

%
%
Note that in the above syntax, the prover is required to generate a witness/proof $\zeta$ for a language $\mathcal{L}_{\st\cmd}$ associated with the command $\cmd$. Concretely, when $\cmd=\scp$, we define:
$$
\mathcal{L}_{\st \scp}
\asn
\{
stm_{\scp}\asn(\vec{o}_{\st \prt}, pp_{\st \prt}, pk_{\st \srv},m)
\;|\;
R_{\st\scp}(stm_{\st\scp},\zeta)=1
\},
$$
and when $\cmd=\ep$, we define: 
$$
\mathcal{L}_{\st\ep}
\asn
\{
stm_{\st\ep}\asn (\vec{g},\vec{pp}^{\st(\textnormal{Evl})}, pk_{\st \srv}, m)
\;|\;
R_{\st\ep}(stm_{\st\ep},\zeta)=1
\}.
$$

Here, $R_{\st \scp}$ is a binary relation asserting that $m$ is a valid solution associated with the single-client puzzle $\vec{o}_{\st \prt}$ under the public data $(pp_{\st \prt},pk_{\st \srv})$, and $R_{\st \ep}$ is a binary relation asserting that $m$ is a valid output associated with the evaluation puzzle $\vec{g}$ under the public data $(\vec{pp}^{\st(\textnormal{Evl})},pk_{\st \srv})$.
Moreover, 
$\vec{pp}^{\st(\textnormal{Evl})}$ denotes the public parameters associated with the evaluated puzzle and includes all public data required to verify the claimed output of the evaluation. 
In particular, for an honest execution with $\cmd=\ep$, the accepted output corresponds to the intended linear combination under the public coefficients, i.e., $m\asn\sum\limits_{\st \prt=1}^{n} q_{\st \prt} \cdot m_{\st \prt}$.
%

%


\subsection{Security Model}\label{sec::Security-Model}


A \vhtlp scheme meets \textit{security} (i.e., privacy and solution-validity), \textit{completeness}, \textit{efficiency}, and \textit{compactness}.  
Each security property is defined via a game between a challenger \chal, representing all honest clients, and an adversary \adv controlling the server. The adversary may additionally obtain the secret keys and local parameters of a subset $\corupt\subseteq C$ of clients. Let \idx be the set of designated leaders. Security requires $|\corupt \cap \idx| \le t$.



\subsubsection{Oracles.} To enable \adv to adaptively choose plaintext solutions and corrupt parties, we provide \adv with access to three oracles $\oracle\asn ( \opgen  , \oeval, \orev )$.
%
%
Below, we define these oracles.


\noindent$\bullet$ $\opgen(st_{\st \chal}, m_{\st \prt}, \Delta_{\st \prt})\rightarrow(\vec{o}_{\st \prt}, pp_{\st \prt})$. It is run by  \chal. 
Upon receiving a query $(m_{\st \prt}, \Delta_{\st \prt})$, where $m_{\st \prt}$ is a plaintext message and $\Delta_{\st \prt}$ is a time parameter,  \chal retrieves $(\kp_{\st \prt}, pk_{\st \srv}, \mxs)$ from its state $st_{\st \chal}$ and  runs $\pgen(m_{\st \prt}, $ $ \kp_{\st \prt}, pk_{\st \srv},$ $ \Delta_{\st \prt}, $ $\mxs)\rightarrow(\vec{o}_{\st \prt}, prm_{\st \prt})$, where $prm_{\st \prt}\asn (sp_{\st \prt}, pp_{\st \prt})$. It returns $(\vec{o}_{\st \prt}, pp_{\st \prt})$ to \adv.

\noindent$\bullet$ $\oeval(\corupt, \idx, t, st_{\st\chal}) \rightarrow
(\vec{g}, \vec{pp}^{\st(\text{Evl})})$. It is run interactively between
the malicious server $\adv$ and  $\chal$. 
Upon invocation, if
$|\corupt \cap \idx| > t$, then $\chal$ returns $(\bot,\bot)$ to $\adv$.
Otherwise, $\chal$ interacts with $\adv$ to run
$
\eval(\langle
\adv,
\prtt_{\st 1}(\Delta,$ $\kp_{\st 1}, prm_{\st 1}, $ $q_{\st 1}, pk_{\st\srv}),
\ldots,$ $
\prtt_{\st n}(\Delta,\kp_{\st n},$ $ prm_{\st n}, $ $ q_{\st n}, $ $pk_{\st\srv})
\rangle)
\rightarrow
(\vec{g}, \vec{pp}^{\st(\text{Evl})}),
$
where the inputs of the honest clients are retrieved by $\chal$ from
$st_{\st\chal}$. Finally, $\chal$ returns $(\vec{g}, \vec{pp}^{\st(\text{Evl})})$ to $\adv$.

%

\noindent$\bullet$ $\orev(\corupt, \idx, t, st_{\st \chal}, \vec{g}, \vec{pp}^{\st(\textnormal{Evl})})
\rightarrow \textsf{trn}^{\st(\textnormal{Evl})}$.
It is run by $\chal$. Upon receiving a query
$(\vec{g}, \vec{pp}^{\st(\textnormal{Evl})})$, where
$(\vec{g}, \vec{pp}^{\st(\textnormal{Evl})})$ is an output pair
previously returned by a specific instance of $\eval$,
if $|\corupt \cap \idx| > t$ or the pair was never generated,
then $\chal$ sets $\textsf{trn}^{\st(\textnormal{Evl})} \asn \emptyset$
and returns it to $\adv$. Otherwise, $\chal$ retrieves from
$st_{\st \chal}$ the messages exchanged between the honest parties
and the clients in $\corupt$ during that specific instance of $\eval$,
sets $\textsf{trn}^{\st(\textnormal{Evl})}$ to this transcript, and
returns $\textsf{trn}^{\st(\textnormal{Evl})}$ to $\adv$.




\subsubsection{Properties.} We formally define \textit{privacy} and \textit{solution validity}, as the two main properties of a $\vhtlp$ scheme. 

Privacy in our setting has two facets. The first is captured by
the experiment $\mathsf{Exp}_{\st \textnormal{prv}}^{\st\adv}$ in Figure~\ref{fig:exp-prv}. It corresponds to the standard model in which the puzzles (corresponding to each client and the evaluation) have not yet been solved. The
second is captured by the experiment $\mathsf{Exp}_{\st \textnormal{prv-eval-sol}}^{\st\adv}$ in Figure~\ref{fig:exp-prv-2} and
corresponds to a stronger and practically relevant setting in which
the puzzle output by $\eval$ is solved, and its solution is revealed,
before the individual client puzzles are solved. This case arises
naturally in asynchronous settings, where the delegated
evaluation may be released earlier than the underlying puzzles. To
the best of our knowledge, prior homomorphic TLP privacy
definitions do not explicitly capture this setting; existing  definitions instead assume simultaneous discovery of
the underlying puzzle solutions.
Formalizing this earlier-revelation setting is nontrivial, since
revealing the evaluation solution would otherwise enable trivial
distinguishing. For this reason,
$\mathsf{Exp}_{\st \textnormal{prv-eval-sol}}^{\st\adv}$ enforces consistency on the challenge vectors for corrupted clients and for the induced public linear combination.


\begin{figure}[h]
\vspace{-2mm}
\centering
{{
\begin{tcolorbox}[
  colback=white,
  colframe=black!50!black,
    fonttitle=\bfseries,
  left=5mm, right=1mm, top=1mm, bottom=1mm,
  boxsep=1mm
]
\vspace{-1mm}
{\small{
$
 \begin{array}{@{}l@{}}
\multicolumn{1}{@{}l@{}}{\hspace{-2mm} \textnormal{Let }\inptCln_{\st \prt}\asn(\Delta,\kp_{\st \prt},prm_{\st \prt}, q_{\st \prt}, pk_{\st \srv})}\\
\multicolumn{1}{@{}l@{}}{\hspace{-2mm}\rule{\linewidth}{0.3pt}}\\%


\setcounter{equation}{0}


\hs\lnum \adv_{\st 1}^{\st \oracle}(1^{\st \lambda}, \tl)\rightarrow pk_{\st \srv}\\
\hs\lnum\textnormal{If}\ \valSetup(pk_{\st \srv})\rightarrow 0, \textnormal{return}\ 0\\

\hs\lnum \mathsf{For}\;  \prt=1,\ldots, n\;\mathsf{do}: \\
\hs\lnum\quad\csetup(1^{\st \lambda})\rightarrow \kp_{\st \prt}:=(sk_{\st \prt}, pk_{\st \prt})\\
\hs\lnum state_{\st 1}\leftarrow \{pk_{\st \srv},  pk_{\st 1},\ldots, pk_{\st n}\}, \corupt\leftarrow \emptyset, \idx\leftarrow \emptyset\\
\hs\lnum K\leftarrow \emptyset, cont\leftarrow True, 
\vec{b}\leftarrow \bf{0}\\
\hs\lnum\label{expr-prv:start-pick-leaders} \textnormal{For } 1,\ldots, \tl \textnormal{ do}: \\
\hs\lnum \quad \textnormal{Select index } a  \textnormal{ from}\ \{1,\ldots, n\} \textnormal{ such that } \advb_{\st a}\notin \idx\\
\hs\lnum\label{expr-prv:end-pick-leaders} \quad \idx \leftarrow \idx \cup\{\advb_{\st a}\}\\
\hs\lnum\label{expr-prv:start-corupt} \textnormal{While } (cont=True) \textnormal{ do}:\\
\hs\lnum  \quad \adv_{\st 1} (state_{\st 1}, K, \idx)\rightarrow (cont, \advb_{\st j})\\
\hs\lnum  \quad \textnormal{If } cont =True, \textnormal{then}\\
\hs\lnum \quad \quad  \hspace{-1mm}\corupt\leftarrow\corupt\  \cup\ \{\advb_{\st j}\}\ \textnormal{\textcolor{NavyBlue}{// $\advb_{\st j}$: a party whose secret state is given to  \adv}} \\
\hs\lnum\label{expr-prv:end-corupt}  \quad \quad \hspace{-1mm} K\leftarrow K \cup\ \{SP_{\st \advb_{\st j}}\}\ \textnormal{\textcolor{NavyBlue}{// $ SP_{\st \advb_{\st j}}$: the secret parameters of $\advb_{\st j}$}}  \\
\hs\lnum\label{expr-prv:pick-messages}  \adv_{\st 1}()\rightarrow (\vec{m} \asn [(m_{\st 0, 1}, m_{\st 1, 1}),\ldots, (m_{\st 0, n}, m_{\st 1, n})], state_{\st 2})\\
\hs\lnum\label{expr-prv:puzzle-gen-start} \textnormal{For }  \prt=1,\ldots, n \textnormal{ do}: \\
\hs\lnum \quad b_{\st \prt}\stackrel{\st \$}\leftarrow\{0, 1\}\\
\hs\lnum \quad \vec{b}[\prt]\leftarrow b_{\st \prt}\\
\hs\lnum\label{expr-prv:puzzle-gen-end}  \quad \pgen(m_{_{\st b_{\prt},\prt}}, \kp_{\st \prt}, pk_{\st \srv}, \Delta_{\st \prt}, \mxs)\hspace{-.8mm} \rightarrow \hspace{-.8mm} (\vec{o}_{\st \prt}, prm_{\st \prt}\asn (sp_{\st \prt}, pp_{\st \prt}))\\
\hs\lnum\label{expr-prv:puzzle-gen-state} \adv_{\st 2}^{\st \oracle}(state_{\st 2}, \vec{o}_{\st 1}, \ldots, \vec{o}_{\st n},  {pp}_{\st 1}, \ldots, {pp}_{\st n})\rightarrow  (b'_{\st \prt}, \prt)  \\
\hs\lnum\label{expr-prv:chalenger-check-single-puzzle} \textnormal{If } (b'_{\st \prt}=\vec{b}[\prt]) \wedge (\prtt_{\st \prt}\notin \corupt), \textnormal{then return } 1\\
\hs \lnum\label{expr-prv:check-threshold} \textnormal{If } |\corupt\cap \idx|>t, \textnormal{then return } 0\\
%
\hs\lnum\label{expr-prv:evaluate}
  \eval(\langle
    \adv_{\st 2},
    {\prtt}_{\st 1}(Inp_{\st 1}),
  \ldots,
    {\prtt}_{\st n}(Inp_{\st n})
  \rangle)\rightarrow(\vec{g}, \vec{pp}^{\st (\text{Evl})}) \\

\hs\lnum\orev(\corupt, \idx, t, st_{\st \chal},  \vec{g}, \vec{pp}^{\st (\text{Evl})})\rightarrow  \textsf{trn}^{\st (\text{Evl})}\\

\hs\lnum\label{expr-prv:reveal}
  \adv_{\st 2}(\vec{g}, \vec{pp}^{\st (\text{Evl})},\textsf{trn}^{\st (\text{Evl})})\rightarrow    (b'_{\st \prt}, \prt)    \\
\hs\lnum\label{expr-prv:2nd-wining-condition}
  \textnormal{If } (b'_{\st \prt}=\vec{b}[\prt]) \wedge (\prtt_{\st \prt}\notin \corupt), \textnormal{then return } 1,
  \textnormal{else return }0
   \end{array}
$
}}
\end{tcolorbox}
\vspace{-4mm}
}}
\caption{Privacy experiment $\mathsf{Exp}_{\st \textnormal{prv}}^{\st\adv}(1^{\st\lambda}, n, \tl, t)$.}\label{fig:exp-prv}
\vspace{-2mm}
\end{figure} 
Each experiment  involves a challenger \chal, representing the honest clients, and a pair of adversaries $\adv\asn(\adv_{\st 1}, \adv_{\st 2})$, as in the standard TLP privacy
definition. We consider a strong adversary who is given access to the three oracles $\oracle\asn ( \opgen  , \oeval, \orev )$ and may additionally obtain the secret keys and local parameters of a subset $\corupt\subseteq C$ of clients chosen adaptively by the adversary.

\noindent\textit{Intuition.} In $\mathsf{Exp}_{\st \textnormal{prv}}^{\st\adv}$, $\adv_{\st 1}$ outputs a pair of challenge messages for each client, the challenger encodes one
randomly chosen message from each pair, and $\adv_{\st 2}$ wins if it
correctly predicts the hidden bit for a non-corrupt client. The game then proceeds through an execution of $\eval$ and transcript leakage through $\orev$, after which $\adv_{\st 2}$ must again predict the hidden bit of a non-corrupt client.


\begin{figure}[h]
\vspace{-1mm}
\centering
{{
\begin{tcolorbox}[
  colback=white,
  colframe=black!50!black,
  fonttitle=\bfseries,
  left=5mm, right=1mm, top=1mm, bottom=1mm,
  boxsep=1mm
]
\vspace{-1mm}
{\small{
$
  \begin{array}{@{}l@{}}

\hspace{-2mm}\textnormal{Let } \inptCln_{\st \prt}\asn(\Delta,\kp_{\st \prt},prm_{\st \prt}, q_{\st \prt}, pk_{\st \srv})\\
\hspace{-2mm}   \textnormal{Lines 1–14 are identical to those of } \mathsf{Exp}_{\st \textnormal{prv}}^{\st\adv}(1^{\st\lambda}\text{, }n\text{, }\tl\text{, }t) \\

\multicolumn{1}{@{}l@{}}{\hspace{-2mm}\rule{\linewidth}{0.3pt}}\\%

\setcounter{equation}{14}

\hs\lnum\label{expr-prv:pick-messages}  \adv_{\st 1}()\rightarrow (\vec{m}^{\st (0)},\vec{m}^{\st (1)}, state_{\st 2})\\

\hs \lnum\label{expr-prv:check-leaked-messages} \textnormal{If } \exists \prt, \textnormal{such that } \prtt_{\st \prt}\in \corupt \textnormal{ and } m^{\st (0)}_{\st \prt}\neq  m^{\st (1)}_{\st \prt} \textnormal{, then return } 0\\

\hs \lnum\label{expr-prv:check-linear-com-condition} \textnormal{If } \sum\limits^{\st n}_{\st\prt=1}
q_{\st \prt}\cdot m^{\st (0)}_{\st \prt}\neq \sum\limits^{\st n}_{\st\prt=1}
q_{\st \prt}\cdot m^{\st (1)}_{\st \prt} \textnormal{, then return } 0 \\

\hs\lnum b\stackrel{\st \$}\leftarrow\{0, 1\}\\

\hs\lnum\label{expr-prv:puzzle-gen-start} \textnormal{For }  \prt=1,\ldots, n \textnormal{ do}: \\
\hs\lnum\label{expr-prv:puzzle-gen-end}  \quad \pgen(m^{\st (b)}_{\st\prt}, \kp_{\st \prt}, pk_{\st \srv}, \Delta_{\st \prt}, \mxs)\rightarrow(\vec{o}_{\st \prt}, prm_{\st \prt}\asn (sp_{\st \prt}, pp_{\st \prt}))\\
\hs \lnum\label{expr-prv:check-threshold} \textnormal{If } |\corupt\cap \idx|>t, \textnormal{then return } 0\\
%
\hs\lnum\label{expr-prv-2:evaluate}
  \eval(\langle
    \adv_{\st 2}^{\st \oracle}(\vec{o}, \vec{pp}, state_{\st 2}), 
    {\prtt}_{\st 1}(\inptCln_{\st 1}),  \ldots,
    {\prtt}_{\st n}(\inptCln_{\st n})
  \rangle)\rightarrow  (\vec{g}, \vec{pp}^{\st (\text{Evl})}) \\

\hs\lnum\orev(\corupt, \idx, t, st_{\st \chal},  \vec{g}, \vec{pp}^{\st (\text{Evl})})\rightarrow  \textsf{trn}^{\st (\text{Evl})}\\

\hs\lnum\label{expr-priv-2:solve} \solv(., ., \vec{g}, \vec{pp}^{\st (\text{Evl})}, \vec{pk}, pk_{\st \srv}, \ep)\rightarrow(m, \zeta)\\

\hs\lnum\label{expr-prv:reveal}
  \adv_{\st 2}(\vec{g}, \vec{pp}^{\st (\text{Evl})},\textsf{trn}^{\st (\text{Evl})}, m, \zeta)\rightarrow b'   \\
\hs\lnum\label{expr-prv:2nd-wining-condition}
  \textnormal{If } b=b', \textnormal{then return } 1,
  \textnormal{else return }0
   \end{array}
$
}}
\end{tcolorbox}
\vspace{-4mm}
}}
\caption{Privacy experiment after post evaluation revelation  $\mathsf{Exp}_{\st \textnormal{prv-eval-sol}}^{\st\adv}(1^{\st\lambda}, n, \tl, t)$.}\label{fig:exp-prv-2}
\vspace{-3mm}
\end{figure} 

In $\mathsf{Exp}_{\st \textnormal{prv-eval-sol}}^{\st\adv}$,
$\adv_{\st 1}$ outputs two message vectors subject to two consistency
constraints: the messages of corrupted clients must be identical in
both vectors, and both vectors must induce the same public linear
combination under the coefficients $q_1,\ldots,q_n$. The challenger
evaluates one of the two vectors at random, reveals the
corresponding evaluation transcript and the solution/proof of the
evaluation puzzle, and $\adv_{\st 2}$ must still guess which vector was
used. Both experiments abort whenever $|\corupt \cap \idx| > t$.

\begin{definition}[Privacy]\label{def:privacy-vh-tlp} 
A $\vhtlp$ scheme is privacy-preserving if, for every security
parameter $\lambda$, every number of clients $n$, every pair of parameters
$\tl,t$ such that $1 \le t < \tl \le n$, every choice of hiding-time
parameter $\Delta^{\st \star} \in \{\Delta,\Delta_{\st 1},\ldots,\Delta_{\st n}\}$
polynomial in $\lambda$, and every polynomial-size adversary
$\adv\asn(\adv_{\st 1},\adv_{\st 2})$ such that $\adv_{\st 1}$ runs in time
$O(poly(T,\lambda))$ and $\adv_{\st 2}$ runs in time $\delta(T)<T$
using at most $\bar{poly}(T)$ parallel processors, where
$T \asn \Delta^{\st \star} \cdot \mxs$ and $\mxs$ is constant in $\lambda$,
there exists a negligible function $\mu$ such that: 
\begin{equation*}
    \begin{split}
\max \Big\{
&\Pr\!\big[\mathsf{Exp}_{\st \textnormal{prv}}^{\st\adv}(1^{\st\lambda}, n , \tl,  t)\to 1\big],\,\\
&\Pr\!\big[\mathsf{Exp}_{\st \textnormal{prv-eval-sol}}^{\st\adv}(1^{\st\lambda}, n , \tl,  t)\to 1\big]
\Big\}
\le \frac{1}{2} + \mu(\lambda).
    \end{split}
\end{equation*}

\end{definition}

Solution validity ensures that a PPT adversary cannot feasibly produce an invalid solution, whether for an individual client's puzzle or one encoding a linear combination of messages, that successfully passes the verification process. This is captured by the experiment $\mathsf{Exp}_{\st \textnormal{val}}^{\st\adv}$ in Figure~\ref{fig:exp-val} which uses the same
corruption phase and the same oracle access. 

\noindent\textit{Intuition.} After $\adv_{\st 1}$ fixes the client messages
and $\chal$ generates the corresponding puzzles, the parties execute $\eval$ subject to the threshold condition $|\corupt \cap \idx| \le t$. The challenger next solves the evaluation puzzle and reveals the
evaluation transcript. Given this information, $\adv_{\st 2}$ attempts to output an accepting but invalid pair $(m',\zeta')$ for the evaluation
puzzle. The experiment then solves each individual client puzzle,
and $\adv_{\st 2}$ analogously attempts to output an accepting but invalid pair for one client puzzle. The adversary wins if $\ver$ accepts
either forgery.



\begin{figure}[h]
\vspace{-1mm}
\centering
{{
\begin{tcolorbox}[
  colback=white,
  colframe=black!50!black,
    fonttitle=\bfseries,
  left=5mm, right=1mm, top=1mm, bottom=1mm,
  boxsep=1mm
]
\vspace{-1mm}
{\small{
$
  \begin{array}{@{}l@{}}
%

\hspace{-2mm} \textnormal{Let } \inptCln_{\st \prt}\asn(\Delta,\kp_{\st \prt},prm_{\st \prt}, q_{\st \prt}, pk_{\st \srv})\\
\hspace{-2mm}   \textnormal{Lines 1–14 are identical to those of } \mathsf{Exp}_{\st \textnormal{prv}}^{\st\adv}(1^{\st\lambda}\text{, }n\text{, }\tl\text{, }t)\\
\multicolumn{1}{@{}l@{}}{\hspace{-2mm}\rule{\linewidth}{0.3pt}}\\%

\setcounter{equation}{14}

\hs\lnum\label{expr-val:pick-messages}  \adv_{\st 1}()\rightarrow (\vec{m}=[m_{\st 1}, \ldots, m_{\st n}], state_{\st 2})\\
\hs\lnum\label{expr-val:start-genPuz} \textnormal{For }  \prt=1,\ldots, n \textnormal{ do}: \\
\hs\lnum\label{expr-val:endd-genPuz} \quad \pgen(m_{\prt}, \kp_{\st \prt}, pk_{\st \srv}, \Delta_{\st \prt}, \mxs)\rightarrow(\vec{o}_{\st \prt}, prm_{\st \prt}\asn (sp_{\st \prt}, pp_{\st \prt}))\\
\hs\lnum\label{expr-val:check-threshold} \textnormal{If } |\corupt\cap \idx|>t, \textnormal{then return } 0\\
\hs\lnum \label{expr-val:evaluate}
  \eval(\langle
    \adv_{\st 2}^{\st \oracle}(\vec{o},\vec{pp}, state_{\st 2}),
    {\prtt}_{\st 1}(\inptCln_{\st 1}),
 \ldots,
    {\prtt}_{\st n}(\inptCln_{\st n})
  \rangle)\rightarrow(\vec{g}, \vec{pp}^{\st (\text{Evl})})\\
\hs\lnum\label{expr-val:solvPul-2} \solv(., ., \vec{g}, \vec{pp}^{\st (\text{Evl})}, \vec{pk}, pk_{\st \srv}, \ep)\rightarrow(m, \zeta)\\
\hs\lnum\orev(\corupt, \idx, t, st_{\st \chal},   \vec{g}, \vec{pp}^{\st (\text{Evl})})\rightarrow  \textsf{trn}^{\st (\text{Evl})}
\\
\hs\lnum\label{expr-val:reveal}
  \adv_{\st 2}( m, \zeta,  \vec{g}, \vec{pp}^{\st (\text{Evl})}, \textsf{trn}^{\st (\text{Evl})}) \rightarrow (m', \zeta') \\
\hs\lnum \label{expr-val:ver-2}
  \textnormal{If } \ver(m', \zeta', .,., \vec{g}, \vec{pp}^{\st (\text{Evl})}, pk_{\st \srv}, \ep)\rightarrow 1\textnormal{, s.t. }\\
  \hspace{-2.1mm} m'\notin \mathcal{L}_{\ep}\textnormal{, then return } 1 \\
\hs\lnum\label{expr-val:start-solvePuz} \textnormal{For }  \prt=1,\ldots, n \textnormal{ do}: \\
\hs\lnum\label{expr-val:end-solvPul}
  \quad \solv(\vec{o}_{\st \prt}, pp_{\st \prt}, ., ., \vec{pk}, pk_{\st \srv}, \scp)\rightarrow(\bar{m}_{\st \prt}, \zeta_{\st \prt})\\
\hs\lnum\label{expr-val:guess-1}
  \adv_{\st 2}\big((\bar{m}_{\st 1}, \zeta_{\st 1}),\ldots, (\bar{m}_{\st \prt}, \zeta_{\st \prt})\big) \rightarrow (m'_{\st \prt}, \zeta'_{\st \prt}, u)\\
\hs\lnum\label{expr-val:ver-1}
  \textnormal{If }
  \ver(m'_{\st \prt}, \zeta'_{\st \prt}, \vec{o}_{\st \prt}, {pp}_{\st \prt}, .,., pk_{\st \srv}, \scp)\rightarrow 1\textnormal{, s.t. }\\
  \hspace{-2.1mm} m'_{\st \prt}\notin \mathcal{L}_{\scp}\textnormal{, then return } 1\\
   \end{array}
$
}}
\end{tcolorbox}
\vspace{-4mm}
}}
\caption{Solution validity experiment  $\mathsf{Exp}_{\st \textnormal{val}}^{\st\adv}(1^{\st\lambda}\text{, }n\text{, }\tl\text{, }t)$.}\label{fig:exp-val}
\vspace{-2mm}
\end{figure}

\begin{definition}[Solution validity]\label{def:validity-vh-tlp}
A \vhtlp scheme preserves solution validity if, for every security
parameter $\lambda$, every number of clients $n$, and every pair of
parameters $\tl,t$ such that $1 \le t < \tl \le n$, every choice of
hiding-time parameter $\Delta^{\st \star} \in \{\Delta,\Delta_{\st 1},\ldots,\Delta_{\st n}\}$
polynomial in $\lambda$, and every polynomial-size adversary
$\adv \asn (\adv_{\st 1},\adv_{\st 2})$ such that $\adv_{\st 1}$ runs in time
$O(poly(T,\lambda))$ and $\adv_{\st 2}$ runs in time $\delta(T)<T$
using at most $\bar{poly}(T)$ parallel processors, where
$T \asn \Delta^{\st \star} \cdot \mxs$ and $\mxs$ is constant in $\lambda$,
there exists a negligible function $\mu$ such that:
$$
\Pr\!\big[\mathsf{Exp}_{\st \textnormal{val}}^{\st\adv}(1^{\st\lambda}, n , \tl,  t)\to 1\big]
\le \mu(\lambda).
$$
\end{definition}



\begin{definition}[Security]\label{def:sec-def-vh-tlp}  A \vhtlp is secure if it satisfies privacy and solution validity as outlined in Definitions \ref{def:privacy-vh-tlp} and \ref{def:validity-vh-tlp}. 

\end{definition}

Informally, \textit{completeness} considers the behavior of the algorithms in the presence of honest parties. It asserts that a correct solution is always obtained by  $\solv $ and $\ver $ returns 1, provided an honestly generated solution.  
Intuitively, \textit{efficiency} states that (1) $\solv $ returns a solution in polynomial time, (2) $\pgen $ generates a puzzle faster than solving it, and (3) the running time of $\eval $ is faster than solving any puzzle involved in the evaluation. Informally, \textit{compactness} requires that the size of evaluated ciphertexts is independent of the complexity of the evaluation function \func. Appendix \ref{sec::Completeness-Definition} defines completeness, efficiency, and compactness. 
%


\section{\tf}

In this section, we present \tf, a protocol that realizes \vhtlp. We must address several challenges to develop an efficient scheme. These challenges and our approaches to tackling them are outlined in Section \ref{sec::Challenges-to-Overcome}.  Subsequently, we explain \tf in Section \ref{sec::Tempora-fusion-protocol}. 


\subsection{Challenges to Overcome}\label{sec::Challenges-to-Overcome}

\subsubsection{Defining an Identical Group for all Puzzles.}

To facilitate correct computation on puzzles (or ciphertexts), the puzzles must be defined over the same group or field. For instance, in the case of the RSA-based time-lock puzzle, the puzzles can be defined over the same group  $\mathbb{Z}_{\st N}$. There are a few approaches to deal with it. Below, we briefly discuss them. 


\noindent(1) \textit{Jointly computing ${ N}$}: With this approach, all clients agree on two large prime numbers $p_{\st 1}$ and $p_{\st 2}$ and then compute $N \asn p_{\st 1}\cdot p_{\st 2}$, where $\log_{\st 2}(N)$ is a security parameter. However, this approach will not be secure if one of the clients reveals the secret key $\phi(N) \asn (p_{\st 1}-1)\cdot (p_{\st 2}-1)$ to the server \srv. In this case, \srv can immediately retrieve the honest party's plaintext message without performing the required sequential work. Another approach is to require all participants (i.e., the server and all clients) to use secure multi-party computation (e.g., the solution proposed in \cite{BonehF97}) to compute $N$, ensuring that no one learns $\phi(N)$. However, this approach requires that all clients are known and interact with each other at the outset of the scheme, which (a) is a strong assumption avoided in time-lock puzzle literature and (b) significantly limits the scheme's flexibility and scalability.  Also, through this approach, it is not clear how each client can efficiently generate its puzzle without the knowledge of $\phi(N)$ and without relying on a trusted setup. 

\noindent(2) \textit{Using a trusted setup}: Through this approach, a fully trusted third party generates $N\asn p_{\st 1}\cdot p_{\st 2}$ and publishes only $N$. As shown in~\cite{MalavoltaT19}, in the trusted setup setting, it is possible to efficiently generate puzzles without knowing  $\phi(N)$. However, fully trusting the third party and assuming that it will not collude with and not reveal the secret to \srv may be considered a strong assumption, and not be desirable in scenarios where the solutions are highly sensitive. The homomorphic TLPs proposed in \cite{MalavoltaT19,dujmovic2023time} rely on a trusted setup. 

\noindent(3) \textit{Using the class group of an imaginary quadratic order}: Through this approach, one can use the class group of an imaginary quadratic order \cite{BuchmannW88}. However, employing this approach will hamper the scheme's efficiency as the puzzle generation phase will no longer be efficient, as highlighted in  \cite{MalavoltaT19}.  Hence, it will violate Requirement~\ref{efficiency-requirement-genpuz} in Definition~\ref{def:efficiency-vh-tlp}, i.e., the efficiency of generating puzzles.


To address this challenge, we propose and use the following new technique, which has been simplified for the sake of presentation. We require the server \srv, only once, to generate and publish a sufficiently large prime number $p$, e.g.,  $\log_{\st 2}(p)\geq 128$. 
We allow each client to \textit{independently} choose its $p_{\st 1}$ and  $p_{\st 2}$ and compute $N\asn p_{\st 1}\cdot p_{\st 2}$. Thus, different clients will have different values of $N$. As in the original RSA-based TLP \cite{Rivest:1996:TPT:888615}, each client generates $a\asn 2^{\st T}\bmod \phi(N)$ for its time parameter $ T$, picks a random value $r$, and then generates  $mk\asn r^{\st a}\bmod N$. 

Now, each client derives two pseudorandom values from $mk$ as: 
 $k\asn\prf(1, mk),$ $ s\asn\prf(2, mk)$,  and then masks its message using the derived values as one-time pads. For instance,  a client's puzzle with solution $m$ is now: 
 $o \asn k\cdot (m+s)\bmod \prm$. 
Now all clients' puzzles are defined over a field, $\mathbb{F}_{\st \prm}$. 
Given each puzzle $o$ (as well as public parameters $N, \prm$, and $r$), a server \srv can find the solution, by computing $mk$ where  $mk\asn r^{\st 2^{\st T}}\bmod N$ through repeated squaring of $r$ modulo $N$, deriving pseudorandom values $k$ and $s$ from $mk$, and  decrypting $o$ to get $m$.








\subsubsection{Supporting Homomorphic Linear Combination.} Establishing a field within which all puzzles are defined, we briefly explain the new techniques we utilize to facilitate a homomorphic linear combination of the puzzles. Recall that each client uses different random values (one-time pads) to mask its message. Therefore, naively applying linear combinations to the puzzles will not yield a correct result. 
To maintain the result's correctness (i.e., completeness), we require the clients to \textit{switch} their blinding factors to new ones when they decide to let \srv find a linear combination of the puzzles. 
To this end, a small subset of the clients (as leaders) independently generates new blinding factors. These blinding factors are generated in such a way that after a certain time, when \srv solves puzzles related to the linear combination, \srv can remove these blinding factors. Each leader sends (the representation of some of) the blinding factors to the rest of the clients.

To switch the blinding factors securely, each client participates in an instance of $\ole^{\st +}$ with \srv. In this case, the client's input is (some of) the new blinding factors and the inverse of the old ones, while the input of \srv is the client's puzzle. $\ole^{\st +}$ returns the output with refreshed blinding factors to \srv. 
To ensure that \srv will learn only the linear combination of honest clients' solutions, without learning a solution for a client's puzzle, we require the leaders to generate and send to the rest of the clients some (of the keys used to generate) zero-sum pseudorandom values such that if all these values are summed up, they will cancel out each other. 
Each client also inserts these pseudorandom values into the instance of $\ole^{\st +}$ that it invokes with \srv, such that the result returned by $\ole^{\st +}$ to \srv will also be blinded by these pseudorandom values.



\subsubsection{Efficient Verification of the Computation.} It is essential for a homomorphic TLP to enable a verifier to check whether the result computed by \srv is correct. However, this is a challenging goal to achieve for several reasons; for instance, (1) each client does not know other clients' solutions, (2) each client has prepared its puzzle independently of other clients, (3) server \srv may exclude or modify some of the clients' puzzles before, during, or after the computation, and (4)  server \srv may gain access to secret parameters of some (leader) clients, which could help it compromise the correctness of the computation.

To achieve the above goal without using costly primitives, we rely on the following techniques. Instead of using plaintext message $m$ as a solution, we represent $m$ as a polynomial ${\pi}(x)$, and use ${\pi}(x)$'s point-value representation to represent $m$. 
We require each leader client to pick a random root and insert it into its outsourced puzzle (which is now a blinded polynomial), during the invocation of   $\ole^{\st +}$ with \srv. It commits to this root (using a random value that \srv can discover when it solves a puzzle related to the linear combination) and publishes the commitment. 
Each leader client sends (a blinded representation of) the root to the rest of the clients that insert it into their outsourced puzzle during the invocation of   $\ole^{\st +}$. Server \srv sums all the outputs of $\ole^{\st +}$ instances and publishes the result. To find the plaintext result, \srv needs to solve a small set of puzzles. We will shortly explain why we are considering a set of puzzles rather than just a single puzzle.

If \srv follows the protocol's instruction, all roots selected by the leader clients will appear in the resulting polynomial that encodes the linear combination of all clients' plaintext solutions. However, if \srv misbehaves, then the honest clients' roots will not appear in the result (according to Theorem \ref{theorem::Unforgeable-Encrypted-Polynomial}). Thus, a verifier can detect it.   
For \srv to prove the computation's correctness,  after it solves the puzzles related to the computation, it removes the blinding factors. Then, it extracts and publishes  (i) the computation result (i.e., the linear combination of the plaintext solutions), (ii) the roots, and (iii) the randomness used for the commitments.    Given this information and public parameters, anyone can check the correctness of the computation's result. 
%
%
%
%
%
%
Now, we outline how we addressed the challenges laid out above.

\setcounter{equation}{0}
\noindent$\bullet$  Challenge ($\countt$): \textit{each client does not know other clients' solutions}: during the invocation of $\ole^{\st +}$ they all agree on and insert certain roots to it, this ensures that all clients' polynomials have the same set of common roots.  Thus, now they know what to expect from a correctly generated result. 

\noindent$\bullet$  Challenge ($\countt$): \textit{each client has prepared its puzzle independently of other clients}:  during the invocation of $\ole^{\st +}$ they switch their old blinding factors to new ones that are consistent with other clients.

\noindent$\bullet$  Challenge ($\countt$): server \srv \textit{may exclude or modify some of the clients' puzzles before, during, or after the computation}: the solutions to Challenges (1) and (2), the support of $\ole^{\st +}$ for verification, and the use of zero-sum blinding factors (that requires \srv sums all outputs of $\ole^{\st +}$ instances) ensure that such misbehavior will be detected by a verifier. Also, requiring \srv to open the commitments for the roots chosen by the leader clients ensures that \srv cannot exclude all parties' inputs,  and come up with its choice of inputs encoding arbitrary roots.

\noindent$\bullet$  Challenge ($\countt$): server \srv \textit{may gain access to some of the (leader) clients' secrets, thereby aiding in compromising the correctness of the computation}: the messages each client receives from leader clients are blinded and reveal no information about their plaintext messages, including the roots. Also, each leader client adds a layer of masking (i.e., blinding factor) to the result. Thus, even if the secret keys of all leaders except one are revealed to the adversary, the adversary cannot find the solution sooner than intended. This is because \srv still needs to solve the puzzle of the honest party to remove the associated blinding factor from the result.

\subsection{The Protocol}\label{sec::Tempora-fusion-protocol}

In this section, we present \tf. We first give a high-level overview and then describe the protocol in detail.



\subsubsection{High-Level Overview.} At this level, we provide a broad perspective on the protocol. Initially, each client generates a puzzle encoding a secret solution (or message) and sends the puzzle to a server \srv. Each client may generate and send its puzzle to \srv at different times. 
At this point, the client may locally delete its plaintext solution, but must retain the secret parameters needed to authorize any subsequent evaluation. 
%
%
%
Upon receiving each puzzle, \srv will work on it to find the solution at a certain time. The time that each client's solution is found can be independent of and different from that of other clients. Along with each solution, \srv generates a proof asserting the correctness of the solution. Anyone can efficiently verify the proof. 

Later, some clients whose puzzles have not been discovered yet, authorize \srv to homomorphically combine their puzzles. The combined puzzles will encode the linear combination of their solutions. The computation result will be discovered by \srv after a certain time, possibly before any of their puzzles are discovered. 
%
%
Eventually, \srv finds the solution for the puzzle encoding the computation result. It also generates proof asserting the correctness of the solution. Anyone can efficiently check this proof by ensuring that the result preserves a certain structure. 

Clients need to be online at two types of stages: (i) to publish their independently generated puzzles, and (ii) throughout each interactive Linear Combination phase (Steps \ref{multi-client:Randomly-selecting-leaders}–\ref{eval:ole-detect}), during which all participating clients must be reachable and retain the secret parameters generated at puzzle-creation time. No client involvement is needed while \srv solves any puzzle, when the evaluation result or any individual solution is released, or for verification, which is public. This brief reachability at authorization time does not reduce our setting to simpler primitives. Regular commitments require each committer to return and reveal at opening time. 
Timed commitments remove this reliance but do not suffice either. With a standard (non-homomorphic) timed commitment alone, committing to the aggregate would require it to be known at authorization time, which is precisely what must remain hidden until its release time. 
Homomorphic timed commitments~\cite{ChvojkaJ23} avoid this, but, as discussed in Section~\ref{Related-Work}, they cannot give the aggregate a shorter release delay than the individual commitments. 
%

This asymmetry in client availability arises naturally in the settings of Section~\ref{sec::intro}. In incident reporting, organizations reconnect briefly once the reporting window closes (precisely when all puzzles to be combined have been received) to authorize the aggregate. Months later, when the result is due, a contributor may have been compromised, ceased operating, or may refuse to cooperate once it anticipates an unfavorable outcome. Likewise, field-deployed sensors are reachable during scheduled synchronization windows but may be destroyed, seized, or unreachable thereafter. In both cases, \tf releases and verifies the aggregate without any client's involvement.

\subsubsection{Detailed Construction.} We proceed to provide a detailed description of  \tf. 

\begin{enumerate}[leftmargin=5.3mm]

\item\label{multi-client::Setup-server} \underline{\textbf{Setup}}. $\ssetup(1^{\st \lambda}, \tl)\rightarrow pk_{\st \srv}.$

The server \srv samples a sufficiently large prime $\prm$, sets $\cor\asn\tl+2$, chooses $X\asn\{x_{\st 1}, \ldots, x_{\st \cor}\}$ such that the $x_{\st i}$ are distinct, non-zero, and outside $U$, and publishes $pk_{\st \srv}\asn(\prm, X)$.






\item\label{multi-client::key-gen} \underline{\textbf{Key Generation}}. $\csetup(1^{\st \lambda})\rightarrow \kp_{\st \prt}.$

Each client $\prtt_{\st\prt}$ samples large primes $(\prm_{\st 1}, \prm_{\st 2})$, sets $N_{\st \prt}\asn\prm_{\st 1}\cdot \prm_{\st 2}$,  
computes Euler's totient function of $N_{\st \prt}$: $\phi(N_{\st \prt})\asn( \prm_{\st 1}-1)\cdot( \prm_{\st 2}-1)$, keeps $sk_{_{\st\prt}}\asn\pnp$, and publishes  $pk_{\st\prt}\asn N_{\st \prt}$.








\item\label{tf-Puzzle-Generation-phase} \underline{\textbf{Puzzle Generation}}. $\pgen(m_{\st \prt}, \kp_{\st \prt}, pk_{\st \srv}, \Delta_{\st \prt}, \mxsqr)\rightarrow(\vec{o}_{\st \prt}, $ \\ $prm_{\st \prt}).$ 
Client $\prtt_{\st \prt}$ first checks whether $\valSetup(pk_{\st s})=1$. If not, it returns $(\bot, \bot)$ and takes no further action. 
%
%
It sets $T_{\st\prt}\asn\Delta_{\st \prt}\cdot \mxsqr$, samples $r_{\st \prt} \stackrel{\st\$}\leftarrow\mathbb{Z}_{\st N_{\st \subprt}}$, computes: 
$$a_{\st \prt} \asn 2^{\st T_{\st\subprt}}\bmod \pnp,\qquad mk_{\st \prt}\asn r^{\st a_{{\st \subprt}}}_{\st \prt}\bmod N_{\st \prt},$$
and derives: 
$$ k_{\st \prt}\asn\prf(1, mk_{\st \prt}),\qquad s_{\st \prt}\asn \prf(2, mk_{\st \prt}).$$

For each $i\in\{1, \ldots, \cor\}$, it sets: 
$$z_{\st i, \prt}\asn\prf(i, k_{\st \prt}), \qquad w_{\st i,\prt}\asn\prf(i, s_{\st \prt}).$$

It encodes $m_{\st \prt}$ as ${\pi}_{\st\prt}(x) \asn x+m_{\st \prt} \bmod \prm$. It evaluates $\pi_{\st\prt}$ on each element of $X$, and blinds each value. $\forall i, 1\leq i\leq \cor:$
 $$\pi_{\st i, \prt}\asn\pi_{\st \prt}(x_{\st i}),\quad o_{\st i,\prt} \asn w_{\st i,\prt}\cdot(\pi_{\st i, \prt} +  z_{\st i,\prt}) \bmod \prm.$$

Finally, it computes $com_{\st \prt} \asn \comcom(m_{\st \prt}, mk_{\st \prt})$, publishes: 
$$\vec{o}_{\st \prt} \asn [o_{\st 1,\prt},\ldots, o_{\st \cor,\prt}], \qquad pp_{\st \prt}\asn(com_{\st \prt}, T_{\st \prt},  r_{\st \prt}, N_{\st \prt}),$$
and outputs 
$prm_{\st {\st \prt}}\asn((k_{\st \prt}, s_{\st \prt})), pp_{\st \prt}).$


\item\label{tf-LinearCombination-phase} \underline{\textbf{Linear Combination}}. $\eval(\langle \srv(\vec{o}, \Delta, \vec{pp}, \vec{pk}, pk_{\st \srv}),$ $c_{\st 1}(\Delta, $ $\kp_{\st {\st 1}}, $\\ $ prm_{\st {\st 1}},$ $ q_{\st 1}, pk_{\st \srv}), \ldots, c_{\st n}(\Delta,  $ $\kp_{\st {\st n}}, prm_{\st {\st n}}, q_{\st n},$  $pk_{\st \srv}) \rangle)\rightarrow(\vec{g}, \vec{pp}^{\st(\text{Evl})})$. 

In this phase, the parties produce messages that allow \srv to learn a linear combination of the clients' messages after time $\Delta$.

\begin{enumerate}

\item\label{multi-client:Randomly-selecting-leaders} \underline{\textit{Randomly selecting leaders}}: all the clients agree on a random key $\hat{r}$ (e.g., one selects a random value and sends to the rest). Each $\prtt_{\st\prt}$ deterministically computes the indices of the $\tl$ leader clients:
$$
\forall j, 1\leq j\leq \tl:\qquad idx_{\st j}\asn\g(j||\hat{r}) .
$$
Let \idx be the set of these $\tl$ clients.

\item\underline{\textit{Leader preprocessing}}: each leader client $\prtt_{\st\prt}\in \idx$ performs the following steps.

\begin{enumerate}

\item\underline{\textit{Generating temporary keys and masks}}: it computes: 
$$
b_{\st \prt}\asn2^{\st Y}\bmod \pnp,\qquad Y=\Delta\cdot \mxsqr,
$$
samples $h_{\st \prt}\stackrel{\st\$}\leftarrow\mathbb{Z}_{\st N_{\st \prt}}$, and sets: 
$$
tk_{\st\prt} \asn h_{\st \prt}^{\st b_{\st\prt}} \bmod N_{\st \prt}.
$$
It then derives: 
$$
\qquad k'_{\st \prt}\asn\prf(1, tk_{\st \prt}),\qquad s'_{\st \prt}\asn\prf(2, tk_{\st \prt}),
$$
and, for every $i$, where $1\leq i\leq \cor:$ 
$$\qquad
w'_{\st i,\prt}\asn\prf(i, s'_{\st \prt}),\qquad
z'_{\st i,\prt}\asn\prf(i, k'_{\st \prt}).
$$

Moreover, for each client $\cl_{\st l}\in C\setminus\prtt_{\st\prt}$, it samples
$f_{\st l}\stackrel{\st\$}\leftarrow\{0,1\}^{\st poly(\lambda)}$ and sends $f_{\st l}$ to $\cl_{\st l}$.

\item\underline{\textit{Generating a blinded root}}: it samples
$\rt_{\st \prt}\stackrel{\st\$}\leftarrow\mathbb{F}^{\st *}_{\st \prm},$
defines 
$\gamma_{\st \prt}(x)\asn x-\rt_{\st \prt}\bmod \prm,$
and, for every $i$, where $1\leq i\leq \cor$, computes: 
$$\qquad
\gamma_{\st i,\prt}\asn \gamma_{\st \prt}(x_{\st i})\bmod \prm,\quad
\gamma'_{\st i,\prt}\asn \gamma_{\st i,\prt}\cdot w'_{\st i,\prt}\bmod \prm.
$$

It sends $\vec{\gamma}'_{\st \prt}\asn[\gamma'_{\st 1,\prt},\ldots,\gamma'_{\st \cor,\prt}]$ to the other clients, computes: 
$$\qquad
com'_{\st \prt}\asn \comcom(\rt_{\st \prt}, tk_{\st \prt}),
$$
and publishes: 
$ {pp}_{\st \prt}^{\st(\text{Evl})}\coloneqq (h_{\st \prt}, com'_{\st \prt}, Y). $
\end{enumerate}
Let $\vec{pp}^{\st(\text{Evl})}$ contain all triples ${pp}_{\st \prt}^{\st(\text{Evl})}$ published by leaders.

\item\underline{\textit{Preparing client-side values}}: each client $\prtt_{\st\prt}\in C$ receives $(\bar f_{\st l}, \vec{\gamma}'_{\st l})$ from every leader $\cl_{\st l}\in \idx\setminus\prtt_{\st\prt}$, regenerates its original blinding factors: $\forall i, 1\leq i\leq \cor:$
$$ \qquad
z_{\st i,\prt}=\prf(i, k_{\st \prt}),\quad
w_{\st i,\prt}=\prf(i, s_{\st \prt}),
$$
and, for every $i$, where $1\leq i\leq \cor$, sets: 
$$
v_{\st i,\prt}\asn \prod\limits_{\st \forall \cl_{\st l}\in \idx}\gamma'_{\st i,\st l}\bmod \prm .
$$

It also sets: 
$$
\delta_{\st i,\prt}\asn
\begin{cases}
z'_{\st i,\prt} & \text{if }\prtt_{\st\prt}\in \idx,\\
0 & \text{otherwise,}
\end{cases}
$$
and
$$\hspace{9mm}
y_{\st i,\prt}\asn
\begin{cases}
- \hspace{-2mm}
\sum\limits_{\st \forall \cl_{\st l}\in C\setminus\prtt_{\st\prt}} \hspace{-2mm}
\prf(i, f_{\st l})
+ \hspace{-2mm}
\sum\limits_{\st \forall \cl_{\st l}\in \idx\setminus\prtt_{\st\prt}}\hspace{-2mm}
\prf(i, \bar f_{\st l})
\bmod \prm & \text{if }\prtt_{\st\prt}\in \idx,\\
\sum\limits_{\st \forall \cl_{\st l}\in \idx}\prf(i, \bar f_{\st l})\bmod \prm
& \text{otherwise.}
\end{cases}
$$

\item\label{eval:ole-detect} \underline{\textit{Re-encoding outsourced puzzle}}: for every client $\prtt_{\st\prt}\in C$ and every $i$, where $1\leq i\leq \cor$, $\prtt_{\st\prt}$ participates in one instance of $\ole^{\st +}$ with \srv. The inputs of $\prtt_{\st\prt}$ are:
\begin{equation*}
\begin{split}\qquad
e_{\st i} &\asn q_{\st \prt}\cdot v_{\st i,\prt}\cdot (w_{\st i,\prt})^{\st -1}\bmod \prm,\\
e'_{\st i} &\asn -(q_{\st \prt}\cdot v_{\st i,\prt}\cdot z_{\st i,\prt})+\delta_{\st i,\prt}+y_{\st i,\prt}\bmod \prm,
\end{split}
\end{equation*}
and the input of \srv is: 
$e''_{\st i}\asn o_{\st i,\prt}.$ 
Hence, the $i$-th instance of $\ole^{\st +}$ returns to \srv: 
\begin{equation*}
\begin{split}\qquad
d_{\st i,\prt}
&\coloneqq e_{\st i}\cdot e''_{\st i}+e'_{\st i}\\
&= q_{\st \prt}\cdot v_{\st i,\prt}\cdot \pi_{\st i,\prt}+\delta_{\st i,\prt}+y_{\st i,\prt}\bmod \prm.
\end{split}
\end{equation*}
If $\prtt_{\st\prt}$ detects misbehavior during any execution of $\ole^{\st +}$, it sends $\bot$ to all parties and halts.

\item\label{multi-client::Computing-encrypted-linear-combination}\underline{\textit{Computing masked linear combination}}: server \srv computes, for every $i$, where $1\leq i\leq \cor:$ 
\begin{equation*}
\begin{split}
\hspace{12mm} g_{\st i}
&\asn \sum\limits_{\st \forall \prtt_{\st\prt}\in C} d_{\st i,\prt}\bmod \prm\\
&=
\left(
\prod\limits_{\st \forall \prtt_{\st\prt}\in \idx}\gamma_{\st i,\prt}\cdot w'_{\st i,\prt}
\right)
\cdot
\left(
\sum\limits_{\st \forall \prtt_{\st\prt}\in C} q_{\st \prt}\cdot \pi_{\st i,\prt}
\right)
+
\sum\limits_{\st \forall \prtt_{\st\prt}\in \idx} z'_{\st i,\prt}
\bmod \prm.
\end{split}
\end{equation*}

The second equality follows from the fact that the values $y_{\st i,\prt}$ cancel out when summed over all clients.

\item\label{publish-encrypted-solutions}\underline{\textit{Disseminating masked result}}: server \srv publishes 
$\vec{g}\asn[g_{\st 1},$ $\ldots,g_{\st \cor}].$

\end{enumerate} 

\item\label{multi-client::Solving-a-Puzzle} \underline{\textbf{Solving a Puzzle}}. $\solv(\vec{o}_{\st \prt}, pp_{\st \prt}, \vec{g}, \vec{pp}^{\st (\text{Evl})}, \vec{pk}, pk_{\st \srv},\cmd)\rightarrow(m, $ $\zeta)$. 
Server \srv proceeds as follows.


\smallskip
\begin{enumerate}[leftmargin=5mm]
\item\label{solving-linear-combination}\hspace{-2mm}. When solving a puzzle related to the linear combination, i.e., when $\cmd=\ep$:
\begin{enumerate}[leftmargin=4.5mm]

\item \underline{\textit{Recovering the leader keys}}: for each $\prtt_{\st\prt}\in \idx$, \srv recovers
$tk_{\st \prt}=h_{\st \prt}^{\st 2^{\st Y}}\bmod N_{\st \prt}$, 
through repeated squaring of $h_{\st \prt}$ modulo $N_{\st \prt}$, where $(h_{\st \prt},Y)\in \vec{pp}^{\st (\text{Evl})}$ and $N_{\st \prt}\in \vec{pk}$. It then derives: 
$$\qquad
k'_{\st \prt}\asn\prf(1, tk_{\st \prt}),\qquad s'_{\st \prt}\asn\prf(2, tk_{\st \prt}).
$$

\item\label{multi-client::server-side-computatoin-removing-pr-vals} \underline{\textit{Unmasking and reconstructing the polynomial}}: for every $i$, where $1\leq i\leq \cor$, \srv computes: 
\begin{equation*}
\begin{split}
\qquad\qquad\quad \theta_{\st i}
&\asn
\Big(\prod\limits_{\st \forall \prtt_{\st\prt}\in \idx}\prf(i,s'_{\st \prt})\Big)^{\st -1}\hspace{-2mm}
\cdot
\Big(g_{\st i}-\hspace{-2.2mm}\sum\limits_{\st \forall \prtt_{\st\prt}\in \idx}\prf(i,k'_{\st \prt})\Big)
\bmod \prm\\
&=
\Big(\prod\limits_{\st \forall \prtt_{\st\prt}\in\idx}\gamma_{\st i,\prt}\Big)
\cdot
\sum\limits_{\st \forall \prtt_{\st\prt}\in C} q_{\st \prt}\cdot \pi_{\st i,\prt}
\bmod \prm .
\end{split}
\end{equation*}
It then interpolates the polynomial $\theta(x)$ from 
$
(x_{\st 1},\theta_{\st 1}),$ $\ldots,(x_{\st \cor},\theta_{\st \cor}).
$
By construction: 
$$
\qquad\qquad\quad \theta(x)=\prod\limits_{\st \forall \prtt_{\st\prt}\in \idx}(x-\rt_{\st \prt})
\cdot
\sum\limits_{\st \forall \prtt_{\st\prt}\in C} q_{\st \prt}\cdot (x+m_{\st \prt})
\bmod \prm .
$$

\item\label{multi-client::server-side-Extracting-the-linear-combination} \underline{\textit{Extracting the result and valid roots}}: let $cons$ be the constant term of $\theta(x)$. Then, it holds that: 
$$\qquad\qquad
cons\asn
\prod\limits_{\st \forall \prtt_{\st\prt}\in \idx}(-\rt_{\st \prt})
\cdot
\sum\limits_{\st \forall \prtt_{\st\prt}\in C} q_{\st \prt}\cdot m_{\st \prt}
\bmod \prm ,
$$

The server \srv computes: 
\begin{equation*}
\begin{split}\qquad
res
&\asn
cons\cdot
\Big(\prod\limits_{\st \forall \prtt_{\st\prt}\in \idx}(-\rt_{\st \prt})\Big)^{\st -1}
\bmod \prm\\
&=
\sum\limits_{\st \forall \prtt_{\st\prt}\in C} q_{\st \prt}\cdot m_{\st \prt}.
\end{split}
\end{equation*}
Next, \srv extracts the roots of $\theta(x)$ and lets $R$ be the resulting set. For each $\prtt_{\st\prt}\in \idx$, it identifies the valid root $\rt_{\st \prt}\in R$ such that: 
$$\qquad
\comver(com'_{\st \prt},(\rt_{\st \prt}, tk_{\st \prt}))=1.
$$

\item\label{publish-lc-proof} \underline{\textit{Publishing the result}}: \srv publishes the solution 
$m\asn res$ 
and the proof
$\zeta\asn\big\{(\rt_{\st \prt}, tk_{\st \prt})\big\}_{\st \forall \prtt_{\st\prt}\in \idx}.
$

\end{enumerate}

\item\label{multi-client::verifying-a-solution-of-single-puzzle-}\hspace{-2mm}. When solving a puzzle of a single client $\prtt_{\st\prt}$, i.e., when $\cmd=\scp$:
\begin{enumerate}

\item\label{multi-client::Finding-secret-keys-} \underline{\textit{Recovering the client keys}}: \srv recovers 
$
mk_{\st \prt}=r_{\st \prt}^{\st 2^{\st T_{\st \prt}}}\bmod N_{\st \prt}
$
through repeated squaring of $r_{\st \prt}$ modulo $N_{\st \prt}$, where $(T_{\st \prt},r_{\st \prt})\in pp_{\st \prt}$. It then derives: 
$$\qquad
k_{\st \prt}=\prf(1,mk_{\st \prt}),\qquad s_{\st \prt}=\prf(2,mk_{\st \prt}).
$$

\item \underline{\textit{Unmasking and reconstructing the polynomial}}: for every $i$, where $1\leq i\leq \cor$, \srv computes: 
$$\qquad\qquad
z_{\st i,\prt}=\prf(i,k_{\st \prt}),\qquad
w_{\st i,\prt}=\prf(i,s_{\st \prt}),
$$
and unblinds: 
$$\qquad
\pi_{\st i,\prt}\asn \big((w_{\st i,\prt})^{\st -1}\cdot o_{\st i,\prt}\big)-z_{\st i,\prt}\bmod \prm.
$$
It then interpolates ${\pi}_{\st \prt}(x)$ from
$
(x_{\st 1},\pi_{\st 1,\prt}),\ldots,(x_{\st \cor},\pi_{\st \cor,\prt}).
$

\item\label{multi-client::Publishing-the-single-pizzle-solution} \underline{\textit{Publishing the solution}}: \srv takes the constant term of ${\pi}_{\st \prt}(x)$ as the plaintext solution $m_{\st \prt}$ and publishes
$$
m\asn m_{\st \prt},\qquad \zeta\asn mk_{\st \prt}.
$$

\end{enumerate}
\end{enumerate}
 
\item\label{multi-client::verification} \underline{\textbf{Verification}}.
$\ver(m, \zeta, ., pp_{\st \prt}, \vec{g}, \vec{pp}^{\st (\text{Evl})}, pk_{\st \srv}, \cmd)\rightarrow \ddot{v}.$

A verifier (which can be any party, not necessarily $\prtt_{\st\prt}\in C$) proceeds as follows.

\begin{enumerate}
\smallskip
\item\noindent\textbf{Case 1 ($\bm\cmd=\ep$):}
\begin{enumerate}[leftmargin=3mm]

\item\label{step-multi-client-verify-opening} \underline{\textit{Checking the openings and reconstructing $\theta$}}:
it first verifies every opening $(\rt_{\st \prt}, tk_{\st \prt})\in \zeta$ against the corresponding commitment in $\vec{pp}^{\st (\text{Evl})}$:
$$
\forall \prtt_{\st\prt}\in \idx:\qquad
\comver\big(com'_{\st \prt},(\rt_{\st \prt}, tk_{\st \prt})\big)\stackrel{\st ?}=1.
$$
If any check fails, it returns $\ddot{v}\asn 0$. Otherwise, for each $\prtt_{\st\prt}\in \idx$, it derives
$$
k'_{\st \prt}\asn\prf(1, tk_{\st \prt}),\qquad
s'_{\st \prt}\asn\prf(2, tk_{\st \prt}),
$$
and, for every $i$, where $1\leq i\leq \cor$, computes
\begin{equation*}
\begin{split}
\qquad\qquad\quad
\theta_{\st i}
&\asn
\Big(\prod\limits_{\st \forall \prtt_{\st\prt}\in \idx}\prf(i, s'_{\st \prt})\Big)^{\st -1}
\cdot
\Big(g_{\st i}-\sum\limits_{\st \forall \prtt_{\st\prt}\in \idx}\prf(i, k'_{\st \prt})\Big)
\bmod \prm\\
&=
\prod\limits_{\st \forall \prtt_{\st\prt}\in\idx}\gamma_{\st i,\prt}
\cdot
\sum\limits_{\st \forall \prtt_{\st\prt}\in C} q_{\st \prt}\cdot \pi_{\st i,\prt}
\bmod \prm.
\end{split}
\end{equation*}
It then interpolates $\theta(x)$ from: 
$
(x_{\st 1},\theta_{\st 1}), \ldots, (x_{\st \cor}, \theta_{\st \cor}). 
$

\item\label{step-multi-client-check-roots} \underline{\textit{Checking the roots and final result}}:
it checks that every opened root is indeed a root of $\theta$, namely
$$
\forall \prtt_{\st\prt}\in \idx:\qquad
\theta(\rt_{\st \prt})\stackrel{\st ?}=0.
$$
Let $cons$ be the constant term of $\theta(x)$. The verifier computes: 
\begin{equation*}
\begin{split}
res'
&\asn
cons\cdot
\Big(\prod\limits_{\st \forall \prtt_{\st\prt}\in \idx}(-\rt_{\st \prt})\Big)^{\st -1}
\bmod \prm\\
&=
\sum\limits_{\st \forall \prtt_{\st\prt}\in C} q_{\st \prt}\cdot m_{\st \prt}.
\end{split}
\end{equation*}
If all root checks pass and $res'\stackrel{\st ?}=m$, it returns $\ddot{v}\asn 1$; otherwise, it returns $\ddot{v}\asn 0$.

\end{enumerate}

\smallskip

\item\noindent\textbf{Case 2 ($\bm\cmd=\scp$).}
The verifier checks whether the opening pair $(m,\zeta)$ matches the commitment in $pp_{\st \prt}$:
$$
\comver\big(com_{\st \prt}, (m,\zeta)\big)\stackrel{\st ?}=1.
$$
If the check passes, it returns $\ddot{v}\asn 1$; otherwise, it returns $\ddot{v}\asn 0$.

\end{enumerate} 
\end{enumerate}

\begin{theorem}\label{theo:security-of-VH-TLP}
If the sequential modular squaring assumption holds, factoring $N$ is a hard problem, \prf is secure, $\ole^{\st +}$ is secure,  and the commitment scheme is hiding and binding, then the protocol presented above is a secure \vhtlp, regarding Definition~\ref{def:sec-def-vh-tlp}. 
\end{theorem}

\begin{proof}[Proof (sketch)]

We prove Theorem~\ref{theo:security-of-VH-TLP} by establishing privacy and solution validity,
as required by Definition~\ref{def:sec-def-vh-tlp}.  We condition on non-aborting executions, so $|\corupt \cap \idx| \le t$.  Since $\tl > t$, at least one leader remains honest. This guarantees that at least one evaluation-side seed of \prf remains hidden until the prescribed delay. For privacy, we use hybrids: we first replace the honest parties' \prf outputs by uniform field elements, and then replace their $\ole^{\st +}$ interactions by simulated transcripts. The indistinguishability of these hybrids follows from the hardness of sequential modular squaring and factoring, \prf security, $\ole^{\st +}$ privacy, and commitment hiding. 
In the experiment $\mathsf{Exp}_{\st \textnormal{prv}}^{\st\adv}$ (Figure~\ref{fig:exp-prv}), this makes the 
adversary's view independent of the challenged honest plaintext. 
In $\mathsf{Exp}_{\st \textnormal{prv-eval-sol}}^{\st\adv}$ (Figure~\ref{fig:exp-prv-2}) after similar replacements discussed above, the output of $\orev$  has the same distribution whether the challenger uses $\vec m^{\st(0)}$ or $\vec m^{\st(1)}$, because of the hardness of sequential modular squaring and factoring, \prf security, $\ole^{\st +}$ privacy, and commitment hiding, and the two vectors agree on all corrupted coordinates. Moreover, both vectors induce the same public linear combination, so the revealed output is the same in the two executions, and hence so is the corresponding proof.

For solution validity $\mathsf{Exp}_{\st \textnormal{val}}^{\st\adv}$ (Figure~\ref{fig:exp-val}), we use a short game sequence. First, any successful forgery whose accepted opening is inconsistent with an honestly published commitment would violate commitment binding. Next, consider a successful forgery for the evaluation puzzle in which the decoded polynomial differs from the honestly generated one but still preserves all honest committed roots. Such a forgery must modify at least one encoded coordinate while keeping every honest root valid, which is negligible by Theorem~\ref{theorem::Unforgeable-Encrypted-Polynomial}. Moreover, if the adversary tampers with an $\ole^{\st +}$ execution involving an honest party, either the $\ole^{\st +}$ verification detects it, or the tampering changes the evaluated encoding in a way that would still have to preserve all honest committed roots to yield an accepting forgery; this is exactly the case ruled out above.
\end{proof} 

We provide a full proof of Theorem \ref{theo:security-of-VH-TLP} in Appendix \ref{sec::VH-TLP-proof}, and prove the completeness, efficiency, and compactness of \tf in Appendix \ref{theo:completeness-efficiency-compactness}. We also provide further discussion on \tf in Appendix~\ref{sec::remarks}. 




\section{Cost Analysis}\label{sec::Cost-Analysis}

We briefly examine the asymptotic costs of \tf before analyzing its
concrete costs. 

\subsection{Asymptotic Cost}
The computation complexity of each non-leader client is $O(\cor \cdot \tl)$, whereas that of each leader client is $O(\cor \cdot (n + \tl))$. 
In particular, across Phases~\ref{tf-Puzzle-Generation-phase} and
\ref{tf-LinearCombination-phase}, a client $\prtt_{\st \prt}$
performs at most four modular exponentiations. In
Phase~\ref{tf-LinearCombination-phase}, each non-leader client invokes $O(\cor \cdot \tl)$ instances of \prf and \cor\ instances of $\ole^{\st +}$; each leader client invokes
$O(\cor \cdot (n+\tl))$ instances of \prf and the same \cor\ instances of $\ole^{\st +}$. The remaining local computation consists of field additions and multiplications, with the same asymptotic costs.

For verifying an evaluation puzzle $(\cmd=\ep)$, the verifier performs
$O(\alpha\cdot\ell)$ \prf evaluations and field operations, followed by
interpolation of a degree-$(\ell+1)$ polynomial and evaluation at
$\ell$ points. Hence, its overall computation complexity is
$O(\ell^2)$. For verifying a single client's puzzle ($\cmd=\scp$), its cost is
independent of $n$ and $\ell$, as it only invokes the verification
algorithm of the commitment scheme.

The computation complexity of the server is at most
$O(\tl^{\st 2}+\tl\cdot(n+Y))$. Specifically, when solving a puzzle related
to the linear combination $(\cmd=\ep)$, the server $\srv$ engages
\cor instances of $\ole^{\st +}$ with each client and performs
$\cor\cdot n$ field additions. During the solving phase, for each
client in $\idx$, $\srv$ performs $Y$ repeated modular squarings and
derives the corresponding temporary keys; it then performs the
operations needed to reconstruct and process the degree-$(\tl+1)$
polynomial, which costs $O(\tl^{\st 2})$. Thus, when $\cmd=\ep$, the
complexity of $\srv$ is $O(\tl^{\st 2}+\tl\cdot(n+Y))$. When solving a puzzle of a single client ($\cmd=\scp$), the complexity of \srv is $O(\cor+T_{\st \prt})$. 
The communication cost of the server and each client is at most $O(\tl\cdot n)$.



\begin{table*}[!h]
\centering
\caption{Phase timing  results for $\tl = 3$. All values are in milliseconds (ms).}
\label{tab:t3-s128-c2048}
\scalebox{1}{
\begin{tabular}{r r r r c r r r r r}
\toprule
 & \multicolumn{1}{c}{\textbf{Setup}} & \multicolumn{1}{c}{\textbf{Key Gen.}} &\multicolumn{1}{c}{\textbf{Pzl. Gen.}} & \multicolumn{1}{c}{\textbf{Linear Comb.}} &\multicolumn{2}{c}{\textbf{Solve}} &\multicolumn{2}{c}{\textbf{Verify }} & \\
\cmidrule(lr){2-2}\cmidrule(lr){3-3}\cmidrule(lr){4-4}\cmidrule(lr){5-5}\cmidrule(lr){6-7}\cmidrule(lr){8-9}
{$n$} & & & & {Client + Server} & {evlPzl} & {cltPzl} & {evlPzl} & {cltPzl} 
\\
\midrule
 10  &0.57 &320.9 &2.7   & 904.9  & 1.8 & 0.25 & 2.3 & 0.02 
 \\
 20  & 0.48 & 408.2 & 5.1   & 961.3  & 1.7 & 0.22 & 1.9 & 0.02 
 \\
 40  & 0.45 & 420.4 & 10.8  & 883.4  & 2.3 & 0.39 & 2.1 & 0.02 
 \\
 50  & 0.70 & 420.5 & 11.7  & 1034.3 & 1.7 & 0.25 & 2.5 & 0.02 
 \\
100  & 1.1 & 404.5 & 28.1  & 870.5  & 1.9 & 0.30 & 2.6 & 0.03 
\\
500  & 0.86 & 443.5 & 167.5 & 1082.9 & 1.8 & 0.28 & 1.1 & 0.02 
\\
\bottomrule
\end{tabular}
}
\end{table*} 
\subsection{Concrete Cost}


%

\subsubsection{Experimental Setup.}

We implemented \tf in \texttt{C++}. The source code is available at~\cite{implementation}. We used the \texttt{GMP} library~\cite{GranlundGMP} for modular multiple precision arithmetic, the \texttt{NTL} library~\cite{NTL} for
 polynomial arithmetic, the \texttt{Crypto++} library~\cite{CryptoPP} for instantiating the \(\prf\) using AES-128~\cite{DBLP:books/crc/KatzLindell2007}, and the BLAKE3 hash function library~\cite{BLAKE3} for instantiating the hash function of the commitment scheme. Experiments were conducted on a system with 32\,GiB RAM and an Intel i7-1365U CPU running Ubuntu 24.04.3 LTS.

We studied the runtime of \tf by varying the number of puzzles \(n\) and fixing the number of leader clients \(\ell\) per iteration. We executed each client specific algorithm in parallel to other clients. Each experiment was repeated 10 times for every value of $n$, and we report the average. We have instantiated $\ole^{\st +}$ using the protocol in Figure~\ref{fig:OLE-plus-protocol}. That protocol invokes an \ole as a subroutine, which we instantiated via oblivious polynomial evaluation (OPE) of Chang and Lu~\cite[p.375]{ChangL01}, setting the polynomial's degree to 1.  This OPE itself relies on a 1-out-of-2 oblivious transfer (OT) that we instantiated using Supersonic OT~\cite{supersonicOT}. We allowed one of the participating clients to play the role of the semi-honest proxy in Supersonic OT.   

We exclude server-side runtime for solving puzzles through sequential squaring as it remains identical among all TLP schemes for a fixed time parameter $T$. Instead, we only measure additional costs imposed by our scheme on the server. Unless stated otherwise, \prm and each RSA modulus $N$ are assigned default bit lengths of 128 and 2048, respectively. On the same machine, for a baseline comparison, we also executed an implementation of the linear homomorphic  TLP (LH-TLP) of Malavolta and Thyagarajan~\cite{MalavoltaT19} using the library in~\cite{HTLP_implementation}.



\subsubsection{Runtime Analysis.}
This section analyzes the concrete runtime of our scheme. Table~\ref{tab:t3-s128-c2048} summarizes the runtime of our scheme for different phases and varying number of puzzles.


\noindent\textbf{Client Runtime.}  The runtime of each leader and non-leader client is about  $565.6$\,ms and $472.5$\,ms respectively at $\tl=3$. The client's runtime is dominated by the linear combination phase that involves $\ole^{\st +}$. 
On average the clients require to stay online for about $1$\,s to complete the linear combination  phase at $\tl = 3$. This time increases by about $1$\,s when $\tl$ changes from $3$ to $10$, as shown in Figure~\ref{fig:grant-client}.

\begin{center}
  \begin{tikzpicture}
    \begin{axis}[
      width=7cm,
      height=4.2cm,
      grid=both,
          major grid style={dashed,gray!30},
    minor grid style={dotted,gray!20},
    mark options={solid},
      xlabel={Number of puzzles $n$},
      ylabel={Online time (ms)},
      xmode=log,
      ymode=log,
      log basis x=10,
    log basis y=10,
      xmin=8, xmax=650,
      ymin=800, ymax=2500,
      xtick={10,20,40,50,100,500},
      ytick={100,1000,1500,2000},
      log ticks with fixed point,
      xtick={10,20,40,50,100,500},
      xticklabels={10,20,40,50,100,500},
      yticklabel style={font=\footnotesize},
      xticklabel style={font=\footnotesize},
      xmin=8, xmax=650,
      legend style={at={(0.5,-0.38)}, anchor=north,
                    font=\footnotesize, legend columns=3},
    ]
      \addplot[color=col3, mark=circle*, mark size=2pt]
        coordinates {(10,904.88)(20,961.27)(40,883.38)(50,1034.33)(100,870.51)(500,1082.89)};
      \addlegendentry{$\tl = 3,\ $ }
      \addplot[color=col5, mark=square*, mark size=2pt]
        coordinates {(10,1305.84)(20,1164.66)(40,1153.06)(50,1096.27)(100,1315.93)(500,1322.86)};
      \addlegendentry{ $\tl=5,\ $ }
      \addplot[color=col10, mark=triangle*, mark size=2pt]
        coordinates {(10,2043.07)(20,1977.48)(40,1991.49)(50,1843.10)(100,2076.18)(500,2101.69)};
      \addlegendentry{$\tl = 10$}
    \end{axis}
  \end{tikzpicture}
  \captionof{figure}{Time that each leader client needs to be online to complete the linear combination phase in \tf. 
  }
  \label{fig:grant-client}
\end{center} 
\noindent\textbf{Server Runtime.}
The runtime of the server's setup and additional steps in puzzle solving remains below $3$\,ms.  During the linear combination phase, the server is involved for $390.6$\,ms at $\tl=3$.



\noindent\textbf{Verifier Runtime.}
Verifier cost is negligible, consistently remains below $3$\,ms. Its runtime is significantly lower than the runtime of each client and the server. 

\noindent\textbf{Comparing with Baseline LH-TLP.} 
Compared to the baseline LH-TLP of Malavolta and Thyagarajan~\cite{MalavoltaT19}, the linear combination in our scheme has a much higher cost when the number of puzzles is smaller than $10^{\st 3}$, as shown in  Figure~\ref{fig:htlp_vs_ours}.

Our experimental results indicate crossover points at approximately $n=22\text{,}000$, $n=28\text{,}000$, and $n=46\text{,}000$ for $\tl=3$, $\tl=5$, and $\tl=10$, respectively. Beyond the corresponding crossover point, \tf outperforms the LH-TLP of~\cite{MalavoltaT19}. Relative to $\tl=3$, the total linear-combination runtime increases by factors of $1.3\times$ and $2.1\times$ for $\tl=5$ and $\tl=10$, respectively.
%
%
We note that this LH-TLP is not functionally equivalent to \tf; below the crossover it is faster, but it cannot verify that a malicious server used the prescribed puzzles and coefficients. Our additional cost stems mainly from the $\ole^{\st +}$ interactions that enable this verifiability.
 

\begin{figure}[!h]
\centering
\begin{tikzpicture}
\begin{axis}[
    width=7cm,
    height=4.5cm,
    xmode=log,
    ymode=log,
    log basis x=10,
    log basis y=10,
    xlabel={Number of puzzles ($n$)},
    ylabel={Runtime (ms)},
     yticklabel style={font=\footnotesize},
      xticklabel style={font=\footnotesize},
    ytick={0.0001, 1, 1000, 100000},
    grid=both,
    major grid style={dashed,gray!30},
    minor grid style={dotted,gray!20},
    mark options={solid},
    legend style={
        at={(0.5,-0.4)},
        anchor=north,
        legend columns=2
    }
]

\addplot[
    color=blue,
    mark=o,
    thick
] coordinates {
    (10,0.554743)
    (100,4.0704)
    (1000,47.814313)
    (10000,427.558371)
    (100000,4864.277056)
    (1000000,55191.5323)
};

\addplot[
    color=red,
    thick,
    mark=dot
] coordinates {
    (10,904.9)
    (100,870.5)
    (1000,1102.36)
    (10000,1223.10)
    (100000,1150.32)
    (1000000, 1421.7)
};
\addplot[
    color=green,
    thick,
    mark=dot
] coordinates {
    (10,1305.8)
    (100,1315.9)
    (1000,1336.1)
    (10000,1357.9)
    (100000,1398.4)
    (1000000,1413.2)
};
\addplot[
    color=black,
    thick,
    mark=dot
] coordinates {
    (10,2043.1)
    (100,2076.2)
    (1000,2102.5)
    (10000,2067.2)
    (100000,2149.7)
    (1000000,2166.5)
};

\legend{
   { \small{LH-TLP of~\cite{MalavoltaT19}}\ },
    {\small Ours ($\tl = 3$)},
    {\small Ours ($\tl = 5$)},
    {\small Ours ($\tl = 10$)}
}

\end{axis}
\end{tikzpicture}
\caption{Total runtime of the linear combination phase in the LH-TLP of~\cite{MalavoltaT19} and \tf. 
}
\label{fig:htlp_vs_ours}
\vspace{-1mm}
\end{figure} 
\noindent{\textbf{Primitive Runtime.}}
Among the primitives employed in \tf, $\ole^{\st +}$ constitutes the dominant computational cost. Empirical measurements indicate that executing a single instance of $\ole^{\st +}$ requires an average of $157.5$\,ms. In contrast, the remaining primitives contribute negligibly to the overall execution time; in particular, $\prf$ requires $0.0087$\,ms, and hash-based commitment requires $0.0058$\,ms.

\section{Conclusion and Future Work}

We presented \tf, the first homomorphic TLP scheme that supports efficient public verification of both single-client puzzle solutions and homomorphic linear combinations. Our construction lets clients generate puzzles independently, later authorize a linear combination with its own release time, and enables any party to verify the released result without trusted setup or costly asymmetric-key proof systems. We introduced new techniques  and showed through our prototype implementation that verification in \tf is lightweight.

An interesting direction for future work is to strengthen
\tf against malicious clients. In this paper, we consider
semi-honest clients and a strong malicious server. Extending the design to also handle malicious clients, without introducing computationally expensive public-key cryptographic machinery, is an important open problem. Another promising direction is to develop post-quantum secure, verifiable homomorphic TLPs. Prior work has studied post-quantum secure TLP constructions, such as the scheme of Lai and Malavolta~\cite{LaiM24}, but existing post-quantum TLPs do not support verifiable homomorphic operations across multiple
puzzles. Extending such constructions in this direction could substantially broaden the applicability of verifiable homomorphic TLPs.



\section{Acknowledgments}\label{sec::ack} 

Aydin Abadi was supported in part by the Kuwait Foundation for the Advancement of Sciences (KFAS) under project code PA24-6TE-2493. Jakub K. Szelag was supported by the Engineering and Physical Sciences Research Council (EPSRC) through a Doctoral Landscape Award studentship. We used ChatGPT and Grammarly to correct grammar and spelling. We thank the anonymous ACM CCS reviewers for their feedback.

\bibliographystyle{ACM-Reference-Format}
\balance
\bibliography{ref}

\appendix


\section{Open Science}\label{sec::OS}
 To support reproducibility and follow-on research, we have publicly released the artifacts underlying this paper. These materials are available through reference \cite{implementation}. They will also be supplied to the Artifact Evaluation (AE) committee after acceptance.


\section{Ethical Considerations} 

This paper presents a cryptographic construction and prototype evaluation only. It does not involve human subjects, personal data, or live-system experimentation. The work aims to improve privacy and verifiability in delayed-disclosure settings.

\section{Proof of Theorem~\ref{theorem::Unforgeable-Encrypted-Polynomial}}%
\label{sec::proof-of-unforgeable-poly}
We model tampering by a modification function  $\modi:\F_{\st \prm}\times G\rightarrow \F_{\st \prm}$ such
that, for each encoded value $u_{\st i}$, the modified value is defined as
$u'_{\st i}=\modi(u_{\st i},g)$, where $g\in G$ is auxiliary input independent of the secret
encoding parameters. We say that a modification occurs if there exists an
index $i$ such that $u'_{\st i}\neq u_{\st i}$.

\begin{proof}
We analyze the probability that the polynomial $\psi(x)$, reconstructed from a modified set of encoded values, preserves the random root $\beta$ of the original polynomial $\theta(x)=\pi(x)\cdot \zeta(x)$, where $\zeta(x)=x-\beta$ and $\beta$ is sampled uniformly at random from $\F_{\st \prm}$.

Consider a modified vector of encoded values $[u'_1,\dots,u'_\pt]$ such that at least one entry differs from the original, i.e., $\exists i^\star\in[\pt]$ with $u'_{i^\star}\neq u_{i^\star}$. From these values, the decoder forms: 
$$
A \asn \Big\{\big(x_1,y'_1\big),\ldots,\big(x_\pt,y'_\pt\big)\Big\},
\qquad
y'_{\st i} \asn w_{\st i}^{-1}\cdot u'_{\st i} - z_{\st i},
$$
and reconstructs $\psi(x)$ as the unique polynomial of degree at most $\pt-1$ consistent with $A$. 
Let:
$$
B \asn \Big\{\big(x_1,y_1\big),\ldots,\big(x_\pt,y_\pt\big)\Big\},
\qquad
y_{\st i} \asn \theta(x_{\st i})=\pi(x_{\st i})\cdot (x_{\st i}-\beta).
$$

Since $\deg(\theta)=d+1<\pt$, the polynomial $\theta(x)$ is uniquely determined by its values on $x_1,\ldots,x_\pt$, and equals its Lagrange interpolation over these points. Let $L_1(x),\ldots,L_\pt(x)$ be the Lagrange basis polynomials for $x_1,\ldots,x_\pt$, namely:
$$
L_{\st i}(x)\asn \prod_{\substack{k=1\\k\neq i}}^{\pt}\frac{x-x_k}{x_{\st i}-x_k}.
$$

Then, it holds that:
$$
\psi(x)=\sum_{i=1}^{\pt} y'_{\st i}\cdot L_{\st i}(x),
\qquad
\theta(x)=\sum_{i=1}^{\pt} y_{\st i}\cdot L_{\st i}(x).
$$

We define the difference polynomial $\Delta(x)$ as:
$$
\Delta(x)\asn \psi(x)-\theta(x)=\sum_{i=1}^{\pt}(y'_{\st i}-y_{\st i})\cdot L_{\st i}(x).
$$

Let $\delta_{\st i}\asn y'_{\st i}-y_{\st i}$. Using the decoding relation $y_{\st i}=w_{\st i}^{-1}\cdot u_{\st i}-z_{\st i}$, we obtain:
$$
\delta_{\st i}
= (w_{\st i}^{\st -1}\cdot u'_{\st i}-z_{\st i})-(w_{\st i}^{\st -1}\cdot u_{\st i}-z_{\st i})
= w_{\st i}^{\st -1}\cdot(u'_{\st i}-u_{\st i}).
$$

Because $w_{\st i} \sample \F^{\st *}_{\st \prm}$, multiplication by $w_{\st i}^{\st -1}$ is injective, hence $w_{\st i}^{\st -1}\cdot(u'_{\st i}-u_{\st i})\neq 0$ unless $u'_{\st i}=u_{\st i}$. However, $u'_{\st i^\star}\neq u_{\st i^\star}$ for some $i^{\st\star}$. Therefore, $\delta_{\st i^\star}\neq 0$. We also have $L_{\st i^\star}(x_{\st i^\star})=1$. It implies that:
$$
\Delta(x_{\st i^\star})=\sum_{\st i=1}^{\st \pt}\delta_{\st i}\cdot L_{\st i}(x_{\st i^\star})=\delta_{\st i^\star}\cdot L_{\st i^\star}(x_{\st i^\star})=\delta_{\st i^\star}\neq 0.
$$

Hence, the polynomial $\Delta(x)$ is not a zero polynomial. Also, $\deg(\Delta)\leq \pt-1$. Note that $\theta(\beta)=\pi(\beta)\cdot(\beta-\beta)=0$. Hence, it holds that: 
\begin{equation}\label{equ::Delta-and-theta-single}
\psi(\beta)=0
\quad\Longleftrightarrow\quad
\Delta(\beta)=\psi(\beta)-\theta(\beta)=0.
\end{equation}

Thus, it suffices to bound the probability that $\Delta(x)$ vanishes at $\beta$. 

 We fix $i\in[\pt]$. Recall $u_{\st i}=w_{\st i}\cdot (y_{\st i}+z_{\st i})$ with $w_{\st i}\sample\F_{\st \prm}^{\st *}$ and $z_{\st i}\sample\F_{\st \prm}$ independent, and $u'_{\st i}=\modi(u_{\st i}, g)$.
 
Fix any value of $y_{\st i}\in\F_{\st \prm}$ and define $t_{\st i} \asn y_{\st i} + z_{\st i}$. Since $z_{\st i}\sample\F_{\st \prm}$ is uniform, $t_{\st i}$ is uniform over $\F_{\st \prm}$ for every fixed $y_{\st i}$. Moreover, $w_{\st i}\sample\F_{\st \prm}^{\st *}$ is uniform and independent of $z_{\st i}$, and hence independent of $t_{\st i}$. Therefore, for every fixed $y_{\st i}$, the pair $(w_{\st i},t_{\st i})$ is distributed in the same way over $\F_{\st \prm}^{\st *}\times\F_{\st \prm}$. 

Using $t_{\st i} = y_{\st i} + z_{\st i}$, we can rewrite the encoding as $u_{\st i}=w_{\st i}\cdot (y_{\st i}+z_{\st i})=w_{\st i}\cdot t_{\st i}$. Since $u'_{\st i}=\modi(u_{\st i},g)$, it follows that $u'_{\st i}=\modi(w_{\st i}\cdot t_{\st i},g)$. Consequently,
$$
\delta_{\st i} \asn w_{\st i}^{\st -1}\cdot(u'_{\st i}-u_{\st i})
= w_{\st i}^{\st -1}\big(\modi(w_{\st i}\cdot t_{\st i},g)-w_{\st i}\cdot t_{\st i}\big),
$$
which is a function of $(w_{\st i},t_{\st i})$, $g$, and the internal randomness of \modi. Hence, the distribution of $\delta_{\st i}$ is the same for all choices of $y_{\st i}$. 
Since $y_{\st i}$ is a function of $\beta$ (given the fixed $x_{\st i}$'s and $\pi(x)$), it follows that:
$$
\Pr\big[\delta_{\st i}=a \mid \beta=b\big]=\Pr\big[\delta_{\st i}=a\big],
$$
for all fixed $a,b\in\F_{\st \prm}$, and hence $\delta_{\st i}$ is independent of $\beta$. 
Moreover, because $\Delta(x)=\sum_{\st i=1}^{\st \pt}\delta_{\st i}\cdot L_{\st i}(x)$ is a function of $(\delta_1,\ldots,\delta_\pt)$ (and the fixed $x_{\st i}$'s), it follows that $\Delta$ is also independent of $\beta$.

Since $\Delta(x)\neq 0$ and $\deg(\Delta)\le \pt-1$, the polynomial $\Delta(x)$ has at most $\pt-1$ roots in $\F_{\st \prm}$. Since $\beta$ is sampled uniformly at random from $\F_{\st \prm}$ and is independent of $\Delta(x)$, it holds that:
$$
\Pr\big[\Delta(\beta)=0\big]\le \frac{\pt-1}{\prm}.
$$

Combining with Relation~\eqref{equ::Delta-and-theta-single}, we conclude that:
$$
\Pr\big[\psi(\beta)=0\big]\le \frac{\pt-1}{\prm}.
$$

That is,
$$
\Pr\big[\psi(x)\text{ has root }\beta\big]\le \frac{\pt-1}{\prm}.
$$
\end{proof}

\section{Commitment Scheme}\label{subsec:commit}

A commitment scheme involves a \emph{sender} and a \emph{receiver}. It also
involves two phases; namely, \emph{commit} and \emph{open}. In the commit
phase, the sender commits to a message $x$ as $com \asn \comcom(x,r)$, that
involves a secret value $r\stackrel{\st\$}\leftarrow \{0,1\}^{\st\lambda}$. At
the end of the commit phase, the commitment ${com}$ is sent to the receiver. In
the open phase, the sender sends the opening $\hat{x}\asn (x, r)$ to the receiver
who verifies its correctness: $\comver({com},\hat{x})\stackrel{\st ?}=1$ and
accepts if the output is $1$.

A commitment scheme must satisfy two properties: (a)~\textit{hiding}: it is
infeasible for an adversary (i.e., the receiver) to learn any information about
the committed message $x$ until the commitment ${com}$ is opened, and
(b)~\textit{binding}: it is infeasible for an adversary (i.e., the sender) to
open a commitment ${com}$ to different values $\hat{x}'\asn(x',r')$ than that was
used in the commit phase, i.e., infeasible to find $\hat{x}'$, \textit{s.t.}
$\comver({com},\hat{x})=\comver({com},\hat{x}')=1$, where $\hat{x}\neq
\hat{x}'$.

There exist efficient commitment schemes both in (a)~the standard model, e.g.,
Pedersen scheme~\cite{Pedersen91}, and (b)~the random oracle model using the
well-known hash-based scheme such that committing is
$com\asn \mathtt{H}(x||r)$ and $\comver({com},\hat{x})$ requires checking
$\mathtt{H}(x||r)\stackrel{\st ?}={com}$, where $\mathtt{H}:\{0,1\}^{\st
*}\rightarrow \{0,1\}^{\st\lambda}$ is a collision-resistant hash function.

\section{The Enhanced OLE's Protocol}\label{apndx:F-OLE-plus}

The enhanced \ole ensures that the receiver cannot learn anything about the
sender's inputs if it sets its input to $0$, i.e., $c=0$. The enhanced \ole's
protocol (denoted by $\ole^{\st +}$) is presented in
Figure~\ref{fig:OLE-plus-protocol}.

\begin{figure}[ht]
\setlength{\fboxsep}{1pt}
\begin{center}
\begin{boxedminipage}{\linewidth}
\begin{small}
\begin{enumerate}
\item Receiver (input $c \in \F$): Pick a random value,
  $r\stackrel{\st\$}\leftarrow \F$, and send
  $(\mathtt{inputS}, (c^{\st -1}, r))$ to the first $\mathcal{F}_{\st\ole}$.

\item Sender (input $a, b \in \F$): Pick a random value,
  $u \stackrel{\st\$}\leftarrow \F$, and send $(\mathtt{inputR}, u)$
  to the first $\mathcal{F}_{\st\ole}$, to learn $t \asn c^{\st -1}\cdot u + r$.
  Send $(\mathtt{inputS},(t + a, b - u))$ to the second $\mathcal{F}_{\st\ole}$.

\item Receiver: Send $(\mathtt{inputR}, c)$ to the second $\mathcal{F}_{\st\ole}$
  and obtain $k \asn (t+a)\cdot c+(b-u)=a\cdot c + b + r\cdot c$. Output $s\asn k - r\cdot c=a\cdot c + b$. 
\end{enumerate}
\end{small}
\end{boxedminipage}
\end{center}
\caption{\small{Enhanced Oblivious Linear Evaluation ($\ole^{\st +}$)~\cite{GhoshN19}.}}
\label{fig:OLE-plus-protocol}
\end{figure}

\section{The Original RSA-Based TLP}\label{sec::RSA-based-TLP}

Below, we restate the original RSA-based time-lock puzzle proposed
in~\cite{Rivest:1996:TPT:888615}.

\begin{enumerate}[leftmargin=5.5mm]
\item \uline{Setup}: $\mathsf{Setup_{\st TLP}}(1^{\st\lambda}, \Delta, \mxsqr)$.
\begin{enumerate}
  \item Pick at random two large prime numbers, $q_{\st 1}$ and $q_{\st 2}$.
    Compute $N=q_{\st 1}\cdot q_{\st 2}$. Compute Euler's totient function of
    $N$ as $\phi(N)\asn(q_{\st 1}-1)\cdot (q_{\st 2}-1)$.
  \item Set $T\asn\mxsqr\cdot \Delta$ as the total number of squarings needed to
    decrypt an encrypted message $m$.
  \item\label{TLP::pick-k} Generate a key for the symmetric-key encryption:\\
    $\mathtt{SKE.keyGen}(1^{\st \lambda})\rightarrow k$.
  \item Choose a uniformly random value $r\stackrel{\st\$}\leftarrow
    \mathbb{Z}^{\st *}_{\st N}$.
  \item Set $a\asn2^{\st T}\bmod \phi(N)$.
  \item Set $pk\asn(N,T,r)$ as the public key and $sk\asn(q_{\st 1},q_{\st 2},a,k)$
    as the secret key.
\end{enumerate}

\item\label{Generate-Puzzle-} \underline{Generate Puzzle}:
  $\mathsf{GenPuzzle_{\st TLP}}(m,pk,sk)$.
\begin{enumerate}
  \item\label{R-TLP::enc-message} Encrypt the message under key $k$:
    $o_{\st 1}\asn \mathtt{SKE.Enc}(k,m)$.
  \item\label{TLP::mask-k} Encrypt the symmetric-key encryption key $k$:
    $o_{\st 2}\asn k+r^{\st a}\bmod N$.
  \item Set ${o}\asn(o_{\st 1}, o_{\st 2})$ as the puzzle and output it.
\end{enumerate}

\item\underline{Solve Puzzle}: $\mathsf{Solve_{\st TLP}}(pk, {o})$.
\begin{enumerate}
  \item\label{R-TLP::find-b} Find $b\asn r^{\st 2^{\st T}}\bmod N$ through
    repeated squaring of $r$ modulo $N$.
  \item\label{R-TLP::dec-key} Decrypt the key's ciphertext:
    $k\asn o_{\st 2}-b\bmod N$.
  \item\label{R-TLP::dec-message} Decrypt the message's ciphertext:
    $m\asn \mathtt{SKE.Dec}(k, o_{\st 1})$. Output the solution, $m$.
\end{enumerate}
\end{enumerate}

The security of the RSA-based TLP relies on the hardness of the factoring
problem, the security of the symmetric key encryption, and the sequential
squaring assumption.

\begin{theorem}\label{theorem::R-LTP-Sec}
Let $N$ be a strong RSA modulus and $\Delta$ be the period within which the
solution stays private. If the sequential squaring holds, factoring $N$ is a
hard problem, and the symmetric-key encryption is semantically secure, then the
RSA-based TLP scheme is a secure TLP.
\end{theorem}

\section{Sequential and Iterated Functions}\label{sec::equential-squering}

The definitions
of sequential and iterated sequential functions in the presence of an adversary
possessing a polynomial number of processors are formally stated
in~\cite{BonehBBF18}. In this section, we restate them. 

\begin{definition}[$(\Delta,\delta(\Delta))$-Sequential function]
For a function $\delta(\Delta)$, time parameter $\Delta$, and security
parameter $\lambda\asn O(\log(|X|))$, $f:X\rightarrow Y$ is a
$(\Delta,\delta(\Delta))$-sequential function if the following conditions hold:
\begin{itemize}
  \item[$\bullet$] There is an algorithm that for all $x\in X$ evaluates $f$
    in parallel time $\Delta$, using $poly(\log(\Delta),\lambda)$ processors.
  \item[$\bullet$] For all adversaries $\mathcal{A}$ executing in parallel time
    strictly less than $\delta(\Delta)$ with $poly(\Delta,\lambda)$ processors:
    $$\Pr\left[y_{\st A}=f(x)\middle| y_{\st A}\stackrel{\st \$}\leftarrow
    \mathcal{A}(\lambda, x), x\stackrel{\st \$}\leftarrow X\right]\leq
    negl(\lambda)$$
    where $\delta(\Delta)=(1-\epsilon)\Delta$ and $\epsilon<1$.
\end{itemize}
\end{definition}

\begin{definition}[Iterated Sequential function]
Let $\beta: X\rightarrow X$ be a $(\Delta,\delta(\Delta))$-sequential
function. A function $f: \mathbb{N}\times X\rightarrow X$ defined as
$f(k,x)=\beta^{\st (k)}(x)=\overbrace{\beta\circ \beta\circ\ldots\circ\beta}^{\st k\text{ Times}}$
is an iterated sequential function, with round function $\beta$, if for all
$k=2^{\st o(\lambda)}$ the function $h:X\rightarrow X$ defined by $h(x)=f(k,x)$
is $(k\Delta,\delta(\Delta))$-sequential.
\end{definition}

\begin{assumption}[Iterated Squaring]\label{assumption::SequentialSquaring}
Let $N$ be a strong RSA modulus, $r$ be a generator of $\mathbb{Z}_{\st N}$,
$\Delta$ be a time parameter, and $T=poly(\Delta,\lambda)$. For any $\mathcal{A}$
as defined above, there is a negligible function $\mu$ such that:
$$\Pr\left[
  \begin{array}{l}
\mathcal{A}(N, r,y) \rightarrow b \\
\hline
r \stackrel{\st \$}\leftarrow \mathbb{Z}_{\st N}, b\stackrel{\st \$}\leftarrow \{0,1\}\\
\text{if}\ b=0,\ y \stackrel{\st \$}\leftarrow \mathbb{Z}_{\st N}\\
\text{else}\ y=r^{\st 2^{\st T}}
\end{array}
\right]\leq \frac{1}{2}+\mu(\lambda)$$
\end{assumption}

\section{Definitions of Completeness, Efficiency, and Compactness}%
\label{sec::Completeness-Definition}

Informally, \textit{completeness} considers the behaviour of the algorithms in
the presence of honest parties. It asserts that a correct solution will always
be retrieved by $\solv$ and $\ver$ will always return 1, given an honestly
generated solution.

\begin{definition}[Completeness]\label{def:completeness-vh-tlp}
A \vhtlp is complete if for any security parameter $\lambda$, any plaintext
input message $m_{\st 1}, \ldots, m_{\st n}$ and coefficient $q_{\st 1},
\ldots, q_{\st n}$ (where each $m_{\st u}$ and $q_{\st u}$ belong to the
plaintext universe $U$), any security parameters $\tl, t$ (where $1\leq \tl,
t\leq n$ and $\tl\geq t$), any difficulty parameter
$T=\Delta_{_{\st l}}\cdot \mxs$, the following conditions are met.

\begin{enumerate}
  \item\label{def::solving-linear-combination} $\solv(\vec{o}_{\st \prt},
    pp_{\st \prt}, ., ., \vec{pk}, pk_{\st \srv}, \cmd)$ always returns
    $m_{\st \prt}$: {\small$$\Pr\left[\begin{array}{l}\ssetup(1^{\st \lambda}, \tl)\rightarrow pk_{\st \srv}\\\csetup(1^{\st \lambda})\rightarrow \kp_{\st \prt}\\\pgen(m_{\st \prt}, \kp_{\st \prt}, pk_{\st \srv}, \Delta_{\st \prt}, \mxs)\rightarrow(\vec{o}_{\st \prt}, prm_{\st \prt})\\\hline\solv(\vec{o}_{\st \prt}, pp_{\st \prt}, ., ., \vec{pk}, pk_{\st \srv}, \cmd)\rightarrow(m_{\st \prt},.)\\\end{array}\right]=1$$}
    where $pp_{\st \prt}\in prm_{\st \prt}$ and $\cmd=\scp$.

  \item\label{def::linearcomb} $\solv(., .,\vec{g}, \vec{pp}^{\st (\text{Evl})}, \vec{pk}, pk_{\st \srv}, \cmd)$ always returns $\sum_{\st \prt=1}^{\st n} q_{\st \prt}\cdot m_{\st \prt}$: {\small$$\Pr\left[\begin{array}{l}\ssetup(1^{\st \lambda}, \tl)\rightarrow pk_{\st \srv}\\\mathsf{For}\ \prt=1,\ldots, n\ \mathsf{do}:\\\quad\csetup(1^{\st \lambda})\rightarrow \kp_{\st \prt}\\\quad\pgen(m_{\st \prt}, \kp_{\st \prt}, pk_{\st \srv}, \Delta_{\st \prt}, \mxs)\rightarrow(\vec{o}_{\st \prt}, prm_{\st \prt})\\\eval\big(\langle \srv(\vec{o}, \Delta,  \vec{pp}, \vec{pk}, pk_{\st \srv}), \\ \prtt_{\st 1}(\Delta, \kp_{\st 1}, prm_{\st 1}, q_{\st 1}, pk_{\st \srv}), \ldots\rangle\big)\rightarrow(\vec{g}, \vec{pp}^{\st (\text{Evl})})\\\hline\solv(., .,\vec{g}, \vec{pp}^{\st (\text{Evl})}, \vec{pk}, pk_{\st \srv}, \cmd)\rightarrow(\sum_{\st \prt=1}^{\st n} q_{\st \prt}\cdot m_{\st \prt},.)\\\end{array}\right]=1$$}
    where $\cmd=\ep$.

  \item\label{def::ver-single-puzzle} $\ver(m_{\st \prt}, \zeta, \vec{o}_{\st \prt}, pp_{\st \prt}, .,., pk_{\st \srv}, \cmd)$ always returns 1, where $\cmd=\scp$.

  \item\label{def::ver-multiple-puzzle} $\ver(m, \zeta, .,., \vec{g}, \vec{pp}^{\st (\text{Evl})}, pk_{\st \srv}, \cmd)$ always returns 1, where $\cmd=\ep$.
\end{enumerate}
\end{definition}

\begin{definition}[Efficiency]\label{def:efficiency-vh-tlp}
A \vhtlp is efficient if the following conditions are satisfied:
\begin{enumerate}[leftmargin=4.8mm]
  \item\label{efficiency-Polynomial-time-solving} The running time of $\solv$
    is upper bounded by $T\cdot poly(\lambda)$.
  \item\label{efficiency-requirement-genpuz} The running time of $\pgen$ is
    upper bounded by $poly'(\log T,$ $ \lambda)$.
  \item\label{efficiency-requirement-Faster-puzzles-evaluation} The running
    time of $\eval$ is upper bounded by $poly''\Big(\log T, \lambda,$ $ \func
    \big((q_{\st 1}, m_{\st 1}),\ldots,(q_{\st n}, m_{\st n})\big)\Big)$.
\end{enumerate}
\end{definition}

\begin{definition}[Compactness]\label{def:compactness-vh-tlp}
A \vhtlp is compact if $\eval$ always outputs a puzzle whose bit-size is
independent of \func's complexity $O(\func)$:
$$||{\vec{g}}||= poly\Big(\lambda, ||\func \big((q_{\st 1}, m_{\st 1}),\ldots,(q_{\st n}, m_{\st n}) \big)||\Big)$$
\end{definition}



\section{Proof of \tf}\label{sec::VH-TLP-proof}

In this section, we prove the security of \tf, i.e., Theorem~\ref{theo:security-of-VH-TLP}.

\begin{proof}
We prove Theorem~\ref{theo:security-of-VH-TLP} by establishing the
two properties required by
Definition~\ref{def:sec-def-vh-tlp}. First,
Lemma~\ref{lemma:VH-TLP-privacy} shows that \tf is privacy-preserving
with respect to Definition~\ref{def:privacy-vh-tlp}. Next,
Lemma~\ref{lemma:VH-TLP-solution-validity} shows that \tf preserves
solution validity with respect to
Definition~\ref{def:validity-vh-tlp}. The theorem then follows
immediately from Definition~\ref{def:sec-def-vh-tlp}.

Throughout this section, we analyze only non-aborting executions of
the experiments in Section~\ref{sec::Security-Model}. By definition, each experiment returns
$0$ whenever $|\corupt \cap \idx| > t$. Hence it suffices to condition on $|\corupt \cap \idx| \le t.$ 
Since $\tl > t$, this implies that at least one leader remains honest.

\begin{lemma}\label{lemma:VH-TLP-privacy}
If the sequential modular squaring assumption holds, the factorization problem is hard,
\prf is secure, $\ole^{\st +}$ is secure privacy-preserving, and the
commitment scheme satisfies the hiding property, then \tf is privacy-preserving,
regarding Definition~\ref{def:privacy-vh-tlp}.
\end{lemma}

\begin{proof}
We prove both inequalities required by Definition~\ref{def:privacy-vh-tlp}.

\smallskip
\noindent\textbf{Part I: privacy in $\mathsf{Exp}_{\st \textnormal{prv}}^{\st\adv}(1^{\st\lambda}, n, \tl, t)$.}
Fix any honest target client $\prtt_{\st \prt}\notin\corupt$. We compare
two neighboring executions that are identical in all respects except that the
challenger encodes $m_{\st 0,\prt}$ in one execution and $m_{\st 1,\prt}$ in
the other. For $\sigma\in\{0,1\}$, let
$\mathsf{View}^{\st H_i}_{\st \adv,\prt,\sigma}$ denote the full view of
$\adv \asn(\adv_{\st 1},\adv_{\st 2})$ in Hybrid $H_i$ when $m_{\st \sigma,\prt}$
is encoded for $\prtt_{\st \prt}$ and all other messages, corruptions, and
public parameters are fixed as in the experiment.

\noindent\textbf{Hybrid $H_0$.}
This is the real execution of
$\mathsf{Exp}_{\st \textnormal{prv}}^{\st\adv}(1^{\st\lambda}, n, \tl, t)$.

\medskip
\noindent\textbf{Hybrid $H_1$.}
For every honest client $\prtt_{\st \prt}\notin\corupt$, replace all outputs of
$\prf$ derived from the hidden keys $mk_{\st \prt}$ and, when $\prtt_{\st \prt}\in\idx$,
from the hidden keys $tk_{\st \prt}$, by uniformly random elements of the field.
This affects the masks used in the client puzzles $\vec{o}_{\st \prt}$ and also
the masks used in the linear-combination phase, including the values appearing
inside 
$\vec{ \gamma}'_{\st \prt}$, 
$\vec{g}$ 
and $\{d_{\st i,\st \prt}\}$.

The difference between $H_0$ and $H_1$ is negligible. Indeed, before the
relevant puzzles are solved, $\adv$ cannot recover any honest $mk_{\st \prt}$ or
$tk_{\st \prt}$ earlier than the prescribed delay without violating the sequential
modular squaring assumption and the hardness of factoring. 
Note that since $\tl > t$, there exists at least one honest leader
$\prtt_{\st h} \in \idx \setminus \corupt$. Therefore, before the evaluation
puzzle is solved, the adversary does not know the corresponding hidden
seed $tk_{\st h}$ and cannot remove all evaluation-side masking layers
from $\vec g$ earlier than the prescribed delay.

Conditioned on the
computational hiding of these seeds, replacing $\prf$ outputs by uniform field
elements is indistinguishable by the security of $\prf$. Hence, for each
$\sigma\in\{0,1\}$,
$$
\mathsf{View}^{\st H_0}_{\st \adv,\prt,\sigma}
\stackrel{c}{\approx}
\mathsf{View}^{\st H_1}_{\st \adv,\prt,\sigma}.
$$

\noindent\textbf{Hybrid $H_2$.}
Replace every transcript of an $\ole^{\st +}$ instance involving an honest
client by the corresponding simulated transcript guaranteed by the privacy of
$\ole^{\st +}$. In particular, the adversary learns no additional information
about the honest sender's inputs beyond the prescribed receiver output, even if
the malicious server inserts $0$ as its own input, due to the privacy-preserving property of $\ole^{\st +}$. 

Again, the difference is negligible. Thus, for each $\sigma\in\{0,1\}$,
$$
\mathsf{View}^{\st H_1}_{\st \adv,\prt,\sigma}
\stackrel{c}{\approx}
\mathsf{View}^{\st H_2}_{\st \adv,\prt,\sigma}.
$$

In Hybrid $H_2$, every honest client's encoded coordinates are hidden by
uniform field masks, every honest $\ole^{\st +}$ transcript is simulated, and
the public parameters $(\mxsqr, N_{\st \prt},\Delta_{\st \prt}, T_{\st \prt},Y,
r_{\st \prt}, h_{\st \prt})$ are independent of the plaintext messages. The published
commitments are also independent of the adversary's bit guess by the hiding
property of the commitment scheme. Therefore, the adversary's view in $H_2$
does not depend on whether $\prtt_{\st \prt}$ encoded $m_{\st 0,\prt}$ or
$m_{\st 1,\prt}$. Hence: 
$$
\mathsf{View}^{\st H_2}_{\st \adv,\prt,0}
\equiv
\mathsf{View}^{\st H_2}_{\st \adv,\prt,1}.
$$

Combining the above hybrids, we obtain: 
$$
\mathsf{View}^{\st H_0}_{\st \adv,\prt,0}
\stackrel{c}{\approx}
\mathsf{View}^{\st H_0}_{\st \adv,\prt,1}.
$$
Therefore, for any PPT adversary,
$$
\Pr\!\left[
\mathsf{Exp}_{\st \textnormal{prv}}^{\st\adv}(1^{\st\lambda}, n, \tl, t)\rightarrow 1
\right]
\le \frac{1}{2}+\mu(\lambda).
$$

\noindent\textbf{Part II: privacy in $\mathsf{Exp}_{\st \textnormal{prv-eval-sol}}^{\st\adv}(1^{\st\lambda}, n, \tl, t)$.}
Let $\vec{m}^{\st (0)}$ and $\vec{m}^{\st (1)}$ be the two challenge vectors
output by $\adv_{\st 1}$ in Figure~\ref{fig:exp-prv-2}. By the rules of the experiment:
(i) for every corrupt client $\prtt_{\st u}\in\corupt$, the two vectors agree
on the $\prt$-th coordinate, and
(ii) both vectors induce the same public linear combination: 
$$
\sum_{\st \prt=1}^{\st n} q_{\st \prt} \cdot m^{\st (0)}_{\st \prt}
=
\sum_{\st \prt=1}^{\st n} q_{\st \prt} \cdot m^{\st (1)}_{\st \prt}.
$$

For $\sigma\in\{0,1\}$, let
$\mathsf{View}^{\st H_i,\textnormal{eval-sol}}_{\st \adv,\sigma}$ denote the
full view of $\adv$ in Hybrid $H_i$ when the challenger uses
$\vec{m}^{\st (\sigma)}$. We use the same three hybrids $H_0,H_1,H_2$ as
above: $H_0$ is the real execution of
$\mathsf{Exp}_{\st \textnormal{prv-eval-sol}}^{\st\adv}$, $H_1$ replaces all
honest $\prf$ outputs by uniform field elements, and $H_2$ replaces all honest
$\ole^{\st +}$ transcripts by simulated ones. By the same reductions as above,
for each $\sigma\in\{0,1\}$,
$$
\mathsf{View}^{\st H_0,\textnormal{eval sol}}_{\st \adv,\sigma} \stackrel{c}{\approx} \mathsf{View}^{\st H_1,\textnormal{eval sol}}_{\st \adv,\sigma} \stackrel{c}{\approx} \mathsf{View}^{\st H_2,\textnormal{eval sol}}_{\st \adv,\sigma}.
$$

It remains to compare the adversary's view in $H_2$ when the challenger uses $\vec m^{\st (0)}$ with its view in $H_2$ when the challenger uses $\vec m^{\st (1)}$.

In Hybrid $H_2$, $\orev$ may return non-public transcript data.
Nevertheless, this transcript does not help distinguish
$\vec m^{\st(0)}$ from $\vec m^{\st(1)}$: the two vectors coincide on
all corrupted coordinates by Figure~\ref{fig:exp-prv-2}, and the honest-party
$\ole^{\st +}$ interactions are replaced by simulated transcripts.
Therefore,
$\textsf{trn}^{\st(\textnormal{Evl})}$ has the same distribution in the
$b=0$ and $b=1$ executions of $H_2$.

Moreover, Figure~\ref{fig:exp-prv-2} reveals the pair $(m,\zeta)$ obtained
by solving the evaluation puzzle. By the consistency condition above, the
revealed plaintext result $m$ is the same in both branches. Since $\solv(\cdot,\cdot,\vec{g},\vec{pp}^{\st (\textnormal{Evl})},
\vec{pk},pk_{\st \srv}, \ep)$ is deterministic, the proof $\zeta$ is also the same in both branches. Therefore,
$$
\mathsf{View}^{\st H_2,\textnormal{eval-sol}}_{\st \adv,0}
\equiv
\mathsf{View}^{\st H_2,\textnormal{eval-sol}}_{\st \adv,1}.
$$

By transitivity,
$$
\mathsf{View}^{\st H_0,\textnormal{eval-sol}}_{\st \adv,0}
\stackrel{c}{\approx}
\mathsf{View}^{\st H_0,\textnormal{eval-sol}}_{\st \adv,1}.
$$
Hence,
$$
\Pr\!\left[
\mathsf{Exp}_{\st \textnormal{prv-eval-sol}}^{\st\adv}
(1^{\st\lambda},n,\tl,t)\rightarrow 1
\right]
\le \frac{1}{2}+\mu(\lambda).
$$

The two inequalities above are exactly those required by
Definition~\ref{def:privacy-vh-tlp}. Therefore, \tf is privacy-preserving.
\unskip\hfill$\blacksquare$
\renewcommand{\qedsymbol}{} 
 %
\end{proof}

\begin{lemma}\label{lemma:VH-TLP-solution-validity}
If the sequential modular squaring assumption holds, factoring $N$ is hard,
\prf is secure, $\ole^{\st +}$ is secure, and the commitment scheme meets the
binding and hiding properties, then \tf preserves solution validity w.r.t.
Definition~\ref{def:validity-vh-tlp}.
\end{lemma}

\begin{proof}
We prove the claim by a short game sequence.

\noindent\textbf{Game $G_0$.}
This is the real experiment
$\mathsf{Exp}_{\st \textnormal{val}}^{\st\adv}(1^{\st\lambda},n,\tl,t)$.

\noindent\textbf{Game $G_1$.}
Game $G_1$ is identical to $G_0$, except that it aborts whenever
$\adv$ outputs an accepting pair whose opening is inconsistent with an honestly
published commitment. Concretely, $G_1$ aborts if either:
(i) $\adv$ outputs $(m'_{\st u},\zeta'_{\st u})$ for some single puzzle and
$\ver$ accepts although $(m'_{\st u},\zeta'_{\st u})$ does not correspond to
the honest commitment $com_{\st u}$; or
(ii) $\adv$ outputs $(m',\zeta')$ for the evaluation puzzle and
$\ver$ accepts although one of the openings $(root_{\st u},tk_{\st u})$
does not correspond to the honest commitment $com'_{\st u}$.

If $\adv$ distinguishes $G_0$ from $G_1$ with non-negligible probability, then
one obtains a non-negligible attack against the binding property of the
commitment scheme. Hence,
$$
\left|
\Pr[G_0=1]-\Pr[G_1=1]
\right|
\le \mu_1(\lambda), 
$$
for a negligible function $\mu_1$.

\noindent\textbf{Game $G_2$.}
This game is identical to $G_1$, except that it additionally aborts whenever
the decoded polynomial associated with the evaluation puzzle differs from the
honestly induced polynomial and nevertheless still contains all honest roots
accepted by the verifier.

By construction, any successful forgery for the evaluation puzzle that survives
$G_1$ must correspond to a non-trivial modification of at least one encoded
coordinate of the evaluated polynomial, yet still preserve every honest committed
root. By Theorem~\ref{theorem::Unforgeable-Encrypted-Polynomial}, this occurs
only with negligible probability. Moreover, if the adversary tampers with an
$\ole^{\st +}$ execution involving an honest party, either the
$\ole^{\st +}$ verification detects it, or the resulting evaluated encoding is a
modified encoding covered by the same theorem. Therefore,
$$
\left|
\Pr[G_1=1]-\Pr[G_2=1]
\right|
\le \mu_2(\lambda),
$$
for a negligible function $\mu_2$.

Finally, in $G_2$ the adversary can win only with negligible
probability, by the binding property of the commitment scheme and
Theorem~\ref{theorem::Unforgeable-Encrypted-Polynomial}. Hence, for
some negligible function $\mu_3$, 
$$
\Pr[G_2=1] \le \mu_3(\lambda).
$$

Combining the three games, we conclude that
$$
\Pr\!\left[
\mathsf{Exp}_{\st \textnormal{val}}^{\st\adv}
(1^{\st\lambda},n,\tl,t)\rightarrow 1
\right]
\le \mu_1(\lambda)+\mu_2(\lambda)+ \mu_3(\lambda),
$$
which is negligible in $\lambda$. Therefore, \tf preserves solution validity.
\unskip\hfill$\blacksquare$
\renewcommand{\qedsymbol}{} 
\end{proof}

We have shown that \tf is privacy-preserving
(regarding Definition~\ref{def:privacy-vh-tlp}) by Lemma~\ref{lemma:VH-TLP-privacy},
and preserves solution validity
(w.r.t.~Definition~\ref{def:validity-vh-tlp}) by
Lemma~\ref{lemma:VH-TLP-solution-validity}. Hence, by
Definition~\ref{def:sec-def-vh-tlp}, \tf is secure. This proves
Theorem~\ref{theo:security-of-VH-TLP}.
\end{proof} 

\section{Proof of Completeness, Efficiency, and Compactness of \tf}%
\label{theo:completeness-efficiency-compactness}

\begin{theorem}\label{theo:completeness-efficiency-compactness-of-tf}
\tf meets completeness, efficiency, and compactness, regarding
Definitions~\ref{def:completeness-vh-tlp}, \ref{def:efficiency-vh-tlp}, and
\ref{def:compactness-vh-tlp}, respectively.
\end{theorem}

\begin{proof} We prove the three required properties separately. Lemma~\ref{lemma::completeness} establishes completeness, Lemma~\ref{lemma::efficiency} establishes efficiency, and Lemma~\ref{lemma::compactness} establishes compactness. 

\begin{lemma}\label{lemma::completeness}
\tf satisfies completeness, regarding Definition~\ref{def:completeness-vh-tlp}.
\end{lemma}

\begin{proof}
We address the four cases in Definition~\ref{def:completeness-vh-tlp}. First, consider Case~\ref{def::solving-linear-combination} for a single
client puzzle. By the correctness of the underlying RSA-based
TLP~\cite{Rivest:1996:TPT:888615}, the server recovers the master key
$mk_{\st \prt}$ from the published puzzle parameters after the prescribed
delay. Since \prf is deterministic, the same $mk_{\st \prt}$ yields the same
keys $k_{\st \prt}, s_{\st \prt}$ and hence the same blinding factors
$z_{\st i,\prt}$ and $w_{\st i,\prt}$ for all $i$. Therefore, the server recovers
each encoded $y$-coordinate as: 
$$
\pi_{\st i,\prt}=w_{\st i,\prt}^{\st -1}\cdot o_{\st i,\prt}-z_{\st i,\prt}\bmod \prm,
$$
interpolates the original polynomial: 
$$
\pi_{\st \prt}(x)=x+m_{\st \prt}\bmod \prm,
$$
and extracts its constant term $m_{\st \prt}$. Hence, \solv returns the
correct solution for a single client puzzle.

Next, consider Case~\ref{def::linearcomb}. Let $\idx$ be the set of leader
clients. For each leader $\prtt_{\st u}\in\idx$, the correctness of the
underlying TLP implies that the server recovers the temporary key
$tk_{\st u}$ from the corresponding public parameters
$pp_{\st u}^{(\textnormal{Evl})}$ after the prescribed delay. By the
determinism of \prf, each recovered $tk_{\st u}$ yields the same
evaluation-side blinding factors that were used by $\prtt_{\st u}$.
Therefore, after removing all $|\idx|$ blinding layers from $\vec{g}$, the
server obtains the point-value representation of a polynomial: 
$$
\theta(x)=
\left(\prod_{\forall \prtt_{\st u}\in \idx}(x-\rt_{\st u})\right)
\cdot
\left(\sum\limits_{\st \forall\prtt_{\st\prt}\in C} q_{\st \prt}\cdot \pi_{\st \prt}(x)\right)
\bmod \prm,
$$
where $\pi_{\st \prt}(x)=x+m_{\st \prt}\bmod \prm$ for every client
$\prtt_{\st \prt}$. Hence, by interpolation, the server recovers $\theta(x)$.
Expanding $\theta(x)$, its constant term is: 
$$
cons=
\left(\prod_{\forall \prtt_{\st u}\in \idx}(-\rt_{\st u})\right)
\cdot
\left(\sum\limits_{\st \forall\prtt_{\st\prt}\in C} q_{\st \prt}\cdot m_{\st \prt}\right)
\bmod \prm.
$$
Thus, the server recovers the linear combination as: 
$$
res=
cons\cdot
\left(\prod_{\forall \prtt_{\st u}\in \idx}(-\rt_{\st u})\right)^{\st -1}
\bmod \prm
=
\sum\limits_{\st \forall\prtt_{\st\prt}\in C} q_{\st \prt}\cdot m_{\st \prt}.
$$
Therefore, \solv returns the correct linear combination.


Finally, Cases~\ref{def::ver-single-puzzle} and
\ref{def::ver-multiple-puzzle} follow from the correctness of the
commitment verification algorithm, the determinism of \prf, and the
correctness of polynomial interpolation. In the single-puzzle case,
the verifier checks the opening $(m_{\st \prt},mk_{\st \prt})$  against the published
commitment. In the linear-combination case, it checks every
opening $(\rt_{\st \prt},tk_{\st \prt})$ against the corresponding commitment
$com'_{\st \prt}$, removes all evaluation-side blinding layers from $\vec{g}$,
interpolates the same polynomial $\theta(x)$, verifies that all
committed roots are roots of $\theta(x)$, and then recovers the
claimed linear combination. Therefore, \ver accepts every honestly
generated solution. 
\renewcommand{\qedsymbol}{}
\end{proof}
\unskip\hfill$\blacksquare$



\begin{lemma}\label{lemma::efficiency}
\tf satisfies efficiency, regarding Definition~\ref{def:efficiency-vh-tlp}.
\end{lemma}

\begin{proof}
We verify the three conditions of Definition~\ref{def:efficiency-vh-tlp}.

Condition~\ref{efficiency-Polynomial-time-solving}.  
For a single-client puzzle, \solv performs the sequential work of the
underlying TLP, namely $T_{\st \prt}=\Delta_{\st \prt}\cdot \mxsqr$ modular
squarings, followed by a polynomial number of \prf evaluations, field
operations, and polynomial interpolation steps. Hence, its running time is
$O(T_{\st \prt}\cdot poly(\lambda))$.

For an evaluation puzzle, \solv recovers the temporary keys of the leader
clients by solving the corresponding evaluation-side TLP instances, each with
time parameter $Y=\Delta\cdot \mxsqr$, and then performs only polynomial-time
post-processing, including \prf evaluations, field operations, interpolation,
and root extraction. Since the number of leaders $\ell$ is polynomially bounded
in $\lambda$, the total running time is $O(T\cdot poly(\lambda))$, where
$T:=\Delta^{\star}\cdot \mxsqr$ as in Definition~\ref{def:privacy-vh-tlp}.
Thus, \solv satisfies
Condition~\ref{efficiency-Polynomial-time-solving}.

Condition~\ref{efficiency-requirement-genpuz}.  
In \pgen, the dominant operations are the computation of
$a_{\st \prt} \asn 2^{T_{\st \prt}}\bmod \phi(N_{\st \prt})$ and
$mk_{\st \prt} \asn r_{\st \prt}^{a_{\st \prt}}\bmod N_{\st \prt}$. These are modular
exponentiations whose exponents have bit-length $O(\log T_{\st \prt})$ and whose
moduli have bit-length polynomial in $\lambda$. The remaining steps of \pgen,
including \prf evaluations, field additions and multiplications, polynomial
evaluation at the public points in $X$, and commitment generation, are all
polynomial in $\lambda$. Therefore, the running time of \pgen is bounded by
$poly'(\log T_{\st \prt},\lambda)$ for some polynomial $poly'$, as required.

Condition~\ref{efficiency-requirement-Faster-puzzles-evaluation}.  
The algorithm \eval performs no sequential work of length $T$. Instead, each
leader computes an evaluation-side temporary key by modular exponentiation with
an exponent of bit-length $O(\log T)$, derives a polynomial number of
pseudorandom values, and participates in $\alpha$ instances of $\ole^{\st +}$. Each
non-leader also participates in $\alpha$ instances of $\ole^{\st +}$ and performs
only polynomially many field operations. The server aggregates the resulting
values component-wise, which is again polynomial-time. Hence, the total running
time of \eval is bounded by
$poly''(\log T,\lambda,\func)$ for some polynomial $poly''$, where the
dependence on $\func$ is through the public coefficient vector
$(q_{\st 1},\ldots,q_{\st n})$ that specifies the linear combination.
Therefore, \eval satisfies
Condition~\ref{efficiency-requirement-Faster-puzzles-evaluation}.

It follows that Tempora-Fusion satisfies Definition~\ref{def:efficiency-vh-tlp}.
\renewcommand{\qedsymbol}{}
\end{proof}
\unskip\hfill$\blacksquare$


\begin{lemma}\label{lemma::compactness}
\tf satisfies compactness, regarding Definition~\ref{def:compactness-vh-tlp}.
\end{lemma}

\begin{proof}
The algorithm $\eval$ outputs a vector of \cor elements as a puzzle
$\vec{g}\asn[g_{\st 1}, \ldots, g_{\st \cor}]$. The bit size of each element
$g_{\st i}$ is $\log_{\st 2}(\prm)$. Thus:
$$||{\vec{g}}||= poly\Big(\log(\prm), ||\func \big((q_{\st 1}, m_{\st 1}),\ldots,(q_{\st n}, m_{\st n}) \big)||\Big),$$
for a fixed polynomial $poly$.
\renewcommand{\qedsymbol}{}
\end{proof}
\unskip\hfill$\blacksquare$

This concludes the proof of Theorem~\ref{theo:completeness-efficiency-compactness-of-tf}.
\end{proof}

\section{Further Discussion on \tf}\label{sec::remarks}

\subsection{The Use of $\ole^{\st +}$}

The $\ole^{\st +}$ ensures that the homomorphic operation can be securely
operated sequentially multiple times, regardless of the distribution of the
input messages. Specifically, one may try a naive approach where each client
directly sends blinding factors to the server. However, this approach is not
secure if the homomorphic linear combination must be computed multiple times,
because the same one-time-pad-like values would be reused, yielding leakage.

\subsection{The Cardinality of $X$}

In \tf, the number of elements in $X$ is \cor for the following reason. Each
client's outsourced polynomial is of degree 1. During Phase~%
\ref{tf-LinearCombination-phase}, this polynomial is multiplied by $\tl$
polynomials of degree 1, each representing a random root. Thus, the resulting
polynomial will have degree $\tl+1$, and $\cor=\tl+2$ $(x,y)$-coordinate
pairs are sufficient to interpolate the polynomial.

\subsection{Flexible Time-Lock Mechanisms}

\tf offers clients the option to apply time-lock mechanisms to their solutions:
some clients may employ time locks, while others may opt for straightforward
masking of their solutions. To mask a message without a time-lock, the
client can use the same masking method used during Phase~%
\ref{tf-Puzzle-Generation-phase} with the only distinction being the omission
of the base $r_{\st \prt}$ publication.

\clearpage






\end{document}